\documentstyle[11pt,epsfig,amssymb]{article}
\newlength{\dinwidth}
\newlength{\dinmargin}
\newcommand {\nn} {\nonumber}
\newcommand {\half} {\frac{1}{2}}
\newcommand {\p} {\prime}
\newcommand {\G} {{\cal G}}
\newcommand{\td}{\nabla}
\renewcommand{\square}{\Box}
\renewcommand{\theequation}{\arabic{equation}}

\begin{document}
\thispagestyle{empty}
\addtocounter{page}{-1}
\begin{flushright}
SNUTP-07/01, WATPHYS-TP 05/01\\
{\tt hep-th/0104227}
\end{flushright}
\vspace*{1.3cm}
\centerline{\LARGE \bf No ghost state of Gauss-Bonnet interaction}
\vskip0.4cm
\centerline{\LARGE \bf in warped backgrounds}
\vspace*{0.8cm}
\centerline{\bf Y. M. Cho ${}^a$, Ishwaree P. Neupane ${}^{a, b}$ {\rm and}
P. S. Wesson ${}^b$}
\vspace*{0.4cm}
\centerline{\it School of Physics \& Center for Theoretical Physics}
\vspace*{0.1cm}
\centerline{\it Seoul National University, Seoul 151-742 Korea ${}^a$}
\vspace*{0.3cm}
\centerline{\it Department of Physics, University of Waterloo}
\vspace*{0.1cm}
\centerline{\it N2L 3G1, Waterloo, ON, Canada ${}^b$}
\vspace*{0.5cm}
%\centerline{\tt ymcho@yongmin.snu.ac.kr, \quad ishwaree@phya.snu.ac.kr,
%\qquad wesson@astro.uwaterloo.ca}
\vspace*{1.0cm}
\centerline{\bf Abstract}
\vspace*{0.4cm}
A general solution to the Einstein field equations with 
Gauss-Bonnet(GB) term in the $AdS_5$ bulk background implies that 
the GB coupling $\alpha$ can take either sign ($+$ or $-$), though a 
positive $\alpha$ will be more meaningful. By considering linearized 
gravity with the GB term in the Randall-Sundrum(RS) a singular $3$-brane 
model, we study the gravitational interactions between matter 
sources localized on the brane. With correctly defined 
boundary conditions on the brane, we find a smooth behavior of 
graviton propagator and hence the zero-mode solution 
as a $4d$ massless graviton localized on the brane with correct momentum and 
tensor structures. The coupling $\alpha$ 
modifies the graviton propagators both on the brane and in the bulk. The 
issue on ghost state 
of the GB term is resolved, and we find that there is no real ghost 
(negative norm) state of the GB term in the RS single brane picture. The 
latter condition leads to a consistency in the coupling between the brane 
matter and the bulk 
gravity. We also elucidate about the possibilities for behavior of a test 
particle on the brane and in the bulk.
\begin{flushleft}
\end{flushleft}

\vspace*{0.6cm}

\baselineskip=18pt

\newpage

\setcounter{equation}{0}

\section{Introduction}
%%%%%%%%%%%%%%%%%%%%%%%%%%%%%%%%%%%%%%%%%%%%%%%%%%%%%%%%%%%%%%%%%%%%%%%
Recently, a considerable effort has been devoted in exploring possible
phenomenological and observable consequences of the brane-world scenario 
in warped backgrounds.
This is mainly after seemingly two alternative directions pioneered by
ArkaniHamed-Dimopoulos-Dvali (ADD)~\cite{ADD}, and
Randall and Sundrum~\cite{RS} in resolving the mass hierarchy
problem~\cite{CBK} (see~\cite{MRK} for a new approach to hierarchy 
problem), and in explaining a small (or vanishing) $4d$ cosmological 
constant without relying on
supersymmetric approach in theories with large (compact and non-compact) extra 
dimensions. Probably a more interesting avenue is the Randall-Sundrum's 
(we refer RS) a singular $3$-brane model with an infinite extra 
dimension~\cite{RS1}(see~\cite{VAR} for early proposals of the non-compact 
Kaluza-Klein(KK) models). The RS proposal~\cite{RS1} involves a conceptually 
fruitful alternative (to the standard KK compactification) 
scheme to trap gravity on a singular $3$-brane
embedded in higher dimensional bulk, where the usual $4$-dimensional
gravity has been realized as the zero-mode spectrum of the
$5$-dimensional theory on the $4d$ boundary. A non-factorizable 
spacetime geometry induced by a gravitating $3$-brane in the 
higher dimensional spacetimes has a number of novel features, 
for example, a probable connection of the brane-world proposal 
to the $AdS/CFT$ correspondences, a description of the strongly 
coupled four-dimensional conformal field theory with 
an uv cut-off at the position of the RS brane, and a realization of 
static $AdS$ domain wall solution in the pure gravitational background. The 
intriguing ideas behind the RS proposals~\cite{RS, RS1}, in particular for 
opening a new door for thinking 
about gravity in extra dimensions, have also led to interesting consequences 
for brane world black hole~\cite{CHR}, new realizations of brane-world KK
reductions in supersymmetry and supergravity theories~\cite{BMC}, and in 
stringy gravity~\cite{KSS}. The static $AdS$ domain wall was previously 
realized as the BPS domain walls of
supergravity theories~\cite{SJR}, while the RS proposals~\cite{RS, RS1} have
further resulted in new work on $AdS$ gravity walls coupled to scalar
interactions~\cite{PKR}, global black $p$-brane \cite{MRK, IMV}, $D3$ 
brane~\cite{BCM} solutions, and embedding of the brane-world 
scenario into a more complete setting of supergravity theories.

The Einstein equations for the linearized perturbation,
$\eta_{\mu\nu}+h_{\mu\nu}(x,y)$, induced by the matter
source on the brane (or without matter) have been the subject of a number of
papers, including [6, 14-22]. Basically, in the RS set-up, the Einstein
gravity can be combined with a warped geometry in the 
bulk $AdS_5$ and a positive
vacuum energy $3$-brane, and the resultant $4d$ metric fluctuations
are described by an attractive delta-function potential generated by the
$AdS$ bulk curvature($\ell$). A single bound state with zero
energy in the eigen-spectrum represents a $4d$ massless graviton, while
other massive Kaluza-Klein modes living in the $5d$ are observed as 
correction to the Newton's law. This is achieved in the RS model 
with different demands than that of the underlying assumptions 
in the conventional compactified KK theories \cite{ACF}. 
In particular, for an infinite fifth dimension $y$, 
the low-energy degrees of freedom do not restrict to zero modes, the continuum
of KK spectrum can exist with no gap, and that the zero-mode wave function 
depends on $y$ non-trivially. The picture is also distinct in
a sense that the coupling of matter to gravity has been realized through
Neumann-type boundary condition on the brane rather than from the Einstein
equations on the $4d$ boundary~\cite{SMS}. Nevertheless, because of a 
fruitful realization of the $4d$ gravity in the extra dimensions, 
and also due to a natural inherence of the $p$-branes 
(which are capable of carrying matter fields) in 
string/M-theory, the RS brane world proposal is appealing and lively.

Given these considerations, a natural approach on a route to realize a
consistent theory is to introduce the higher curvature terms to the action,
for the effective $4$-dimensional gravity on the $3$-brane should be that of
Einstein plus the higher order curvature corrections . These correction
terms should arise from the low energy effective action of string theory 
or/and as the $1/N$ corrections in the large $N$ limit of some gauge 
theory, and one could introduce
them as a ghost-free Gauss-Bonnet (GB) combination [26-32]. Explicit brane 
solutions for arbitrary order of higher curvature terms with or without 
cosmological constant were also discussed 
in~\cite{Marek}. 
An important aspect concerns whether the higher-curvature terms (in GB 
combination) in the bulk action can result in a localized $4d$ gravity 
or/and still reproduce the correct zero-mode behaviors (both the momentum 
and tensor structure) at long distance scales on the
brane. Our answer to this is positive (see Ref.~\cite{IPN4} for very recent 
results on the effective four-dimensional gravity localized on the 
(intersection of) $n$ domain walls in $(4+n)$-dimensional space-time with 
higher curvature terms). It is known that the generic form of the 
higher curvature terms can delocalize gravity~\cite{CZK} on the brane, and
furthermore, in warped backgrounds with
finite volume non-compact extra-dimension(s), one should be cautious about
such terms due to the possible excitations of unphysical scalar modes (e.g.,
graviscalar or tachyonic mode) and, massless or massive ghosts. But the GB
term does not excite any ghosts in the brane background.
It would appear that in the RS single brane model with positive 
brane tension $3$-brane. In the latter case, a non-trivial GB 
coupling, however, renormalizes the $4d$ Newton constant on 
the brane and also modifies the graviton propagators both on the brane 
and in the bulk, unlike the case is in a flat space-time 
background~\cite{ZWI}.

In this paper, by introducing GB term into the Einstein-Hilbert action, we 
mainly work with the transverse-traceless (TT) components of the metric 
fluctuations in the presence of matter source on the brane. However, one 
should take a proper account of non-TT components of the metric fluctuations 
to derive the correct Neumann boundary condition (Israel junction condition 
and enforcing $Z_2$ symmetry) across the brane and hence an account that of 
brane-shift function ($\hat{\xi}\,^5$) for a matter-localized 
$3$-brane. In such a case an extra polarization due to the trace part of 
the perturbation is actually compensated by the brane-shift function. 
A detailed 
formalism was developed in \cite{GTT, GKR} and analyzed in a
very similar fashion in~\cite{KL} by adding GB term for the non-TT 
components, but in the last context, the theory suffered from a ghost/
extra polarization state of GB term. In various steps our treatments also
follow parallel to~\cite{GKR}. Instead, here we offer some resolution
(or better, reformulation) on the linearized Einstein-Gauss-Bonnet gravity,
in particular the negative norm state of GB term, and find that there is no 
any real ghost state due to such interaction term in the RS 
single brane model.

In Section 2 we present some important features of the EGB theory in an 
$AdS_5$. In Section $3$ we analyse the $5d$ killing equations. 
Sections $4$ and $5$ deal with the general behaviors and stability 
analysis of the graviton propagators, which present the most relevant 
generalization of the RS gravity with the GB term at the linearized 
level. Section $6$ elucidates upon the geodesics in the 
general $AdS_5$ backgrounds. Section $7$ is a conclusion.      
%%%%%%%%%%%%%%%%%%%%%%%%%%%%%%%%%%%%%%%%%%%%%%%%%%%%%%%%%%%%%%%%
\section{Effective action and general solution}
We study the gravitational interactions between matter fields localized on 
a singular $3$-brane model with the Gauss-Bonnet term by considering the 
following effective action defined on the $D$-dimensional
space-time $(M)$, where $\partial M$ represents the $(D-1)$-dimensional 
boundary,
\begin{eqnarray}
S&=&\int_{M} d^D x\,\sqrt{-g}\,\Big\{\kappa^{-1}\big( R-2\Lambda\big)
+\alpha \big( R^2- 4 R_{ab}R^{ab}
+ R_{abcd}R^{abcd}\big)\Big\}\nn\\
&&+2\int_{\partial M}d^d\,x\sqrt{-\gamma}\,\big({\cal L}_m^{bdry}-\sigma(z_i)
\big) +\int_{M} d^{d+1} x\, \sqrt{-g}\,{\cal L}_{m}^{bulk}
\label{action}\,.
\end{eqnarray}
We follow the metric signature $(-,+,\cdots +)$. 
Here $a,\cdots d =0,1,\cdots D$ and $x^a=(x^{\mu},z^i)$, 
where $x^{\mu}(\mu = 0,...,3)$ are the brane coordinates and $z_i$ are the 
bulk coordinates transverse to the brane. $\Lambda$ is the bulk cosmological 
constant in $AdS_D$,
$\sigma(z_i)$ are the brane tensions or vacuum energy of the branes (or an 
interaction thereof), and $\gamma_{\mu\nu}$ is the induced metric on the brane.
As in this paper we work only in the five space-time dimensions, it is useful 
to define the above parameters in $D=5$. The parameters therefore have 
dimensions $[\Lambda]=M^2,\, [\sigma]=M^4,\,
[\alpha]=M$. Since the $D(=d+1)$-dimensional mass is defined by
$\kappa_{d+1} = 16 \pi G_{d+1} = M_*^{1-d}$, 
where $M_*$ is the $(d+1)$-dimensional fundamental mass scale, one can write 
$\kappa_5= M_*^{-3}$. By the same token, since the GB coupling $\alpha$ has
the mass dimension of $M_*^{d-3}$, in five dimensions, 
we define $\alpha=M_* \alpha'$, where $\alpha'$ is the effective 
(dimensionless) GB coupling constant.

The graviton equations derived by varying the above action with respect to
$g^{ab}$ take the form
\begin{equation}
G_{ab}+ \kappa\,H_{ab}= T^{(0)}_{ab}+\frac{\kappa}{2}\,
T^{(m)}_{ab}\,,\label{maineqn}
\end{equation}
where $G_{ab}=R_{ab}-g_{ab}R/2$ and
$H_{ab}$, an
analogue of the Einstein tensor stemmed from the GB term, is given by
\begin{eqnarray}
H_{ab}&=&-\frac{\alpha}{2}\,g_{ab}\,
(R^2-4 R_{cd}R^{cd}+ R_{cdef}R^{cdef})\nn\\
&&+2\alpha \big[ R R_{ab}- 2 R_{acbd}R^{cd} +
 R_{acde}R_b\,^{cde}-2R_a\,^c R_{bc}\big]\,.
\end{eqnarray}
 and,
\begin{eqnarray}
T^{(m)}_{ab}&=&-\frac{2}{\sqrt{-g}}\frac{\delta}{\delta g^{ab}}\int d^{d+1} x
\sqrt{-g}{}{\cal L}_m,\label{tabmatter}\\
T^{(0)}_{ab} &=& -\Lambda g_{ab}-\frac{\sqrt{-\gamma}}
{\sqrt{-g}}\,\gamma_{\mu\nu}\,\delta_a^\mu\,\delta_b^\nu\,\delta(z)\,
\sigma(z)\,.\label{tabzero}
\end{eqnarray}

We are interested in the solutions with a warped metric having the form
\begin{equation}
ds^2= e^{-2A(z)}(\eta_{\mu\nu}dx^{\mu}dx^{\nu} + dz^2)
={\bar g}_{ab}dx^a dx^b\,.
\label{conmetric}
\end{equation}
The system of equations of motion given by~(\ref{maineqn}) has a solution if
one fine-tunes the brane tension to the bulk vacuum energy as 
\footnote{This is merely a reflection of the assumption
that the $3$-brane world volume is Minkowskian.
Under the axial gauge $h_{ai}=0$ and $4d$ 
transverse-traceless (TT) gauge $h_\mu^\mu=0=\partial^m h_{\mu\nu}$, one may 
define $\delta T^{(0)}_{\mu\nu}=\bar{T}_{\mu\lambda} h_\nu^\lambda$.} 
$\Lambda= - 6\kappa_5^{-1}/\ell^2$ and $\sigma(z) = 6 \kappa^{-1}/\ell$, 
whose solution in five space-time dimensions is given by
\begin{equation}
A(z)=\ln(|z|/\ell+1)
\end{equation}
where,
\begin{equation}
\ell^2=\frac{4\alpha'}{M_*^2}\bigg[1\pm
\sqrt{1+\frac{4\alpha'\Lambda}{3M_*^2}}\,\bigg]^{-1}=\ell_{\pm}^2\,.
\label{alsquare}
\end{equation}
To recover the usual Einstein gravity in the RS set-up, the curvature scale
$\ell$ could be set in the order of (or larger than) the fundamental 
(string) scale. From the above relation, we define
$\gamma=4 \alpha\,\kappa\,\ell^{-2}=4\alpha'\,\ell^{-2} M_*^{-2}$ for a 
future use, and pause for a while to regard whether $\gamma$ should be large
or small.

The value of $\gamma$ can be partially fixed from the ratio of $M_{pl}$ and
$M_*$. Specifically, in the presence of Gauss-Bonnet term, $\ell$ admits two
values implied by Eq.(\ref{alsquare})
\begin{equation}
1-\gamma=1-4\alpha'\ell^{-2}M_*^{-2}=\mp\sqrt{1+\frac{4\alpha'\Lambda}
{3M_*^2}}\,,\label{limitongamma}
\end{equation}
for $\ell_+$ and $\ell_-$ solutions respectively. For
a large mass hierarchy between $M_*$ and $M_{pl}$, one needs $\ell M_*>>1$.
In the RS scenario, since $\ell M_*\sim 1$, $\gamma$ would be in the order of
$\alpha'$, which is small enough. While, in the
ADD picture, where the fundamental mass scale could arise in $TeV$ range and 
$\ell M_*>>1$ is expected, $\gamma$ would be much smaller than unity. 
The $\ell_+$ solution may violate the weak energy condition for localized 
gravity, the latter condition reads as $2\ell^{-1}\,(1-\gamma/3)>0$ at the 
brane. Though this condition loosely restricts $\gamma$ as $\gamma< 3$, the 
actual graviton propagator analysis reveals that we also 
need to impose $(1-\gamma)>0$ for a definite positive contribution to the 
Newtonian potential from the KK kernel. Further, in order to 
avoid anti-gravity effect ( i.e., $G_4<0$)\footnote{In fact, $G_4\geq 0$ 
limit corresponds to $(1+\gamma)\geq 0$}, one must take $\gamma>-1$ limit 
as well. Hence the effective limits for $\gamma$ become $1>\gamma>-1$ 
(see below). For any values of $\alpha'$ and $\Lambda$, the 
$\ell_+$ solution implies that $\gamma>1$, and hence the leading order 
correction term to the Newtonian potential may appear with a negative sign, 
which may excite ghost states in the background. As seen 
from~(\ref{alsquare}), in the 
$AdS$ background $(\Lambda<0)$, either sign of $\alpha'$ is allowed for
$\ell_-$ solution, but a positive $\alpha'$ would be a better choice, 
for $\alpha'<0$ solution can violate unitarity in some parameter space of 
the full bulk solution. We also note that a de-Sitter bulk solution 
($\Lambda>0$) is not allowed with GB term, because in this case one always
encounters either an anti-gravity effect or finds an imaginary curvature scale.
If one demands $\alpha'<0$ in an $AdS$ background, then the
$\ell_-$ solution may be allowed, admitting negative $\gamma$ but 
satisfying the limit $\gamma >-1$. Indeed, the effective limits 
$1>\gamma>-1$ translate, in terms of bulk parameters, to 
$-3/4<\alpha'\Lambda M_*^{-2}<9/4$, and one then has to satisfy these limits 
with a negative bulk cosmological constant. 
%%%%%%%%%%%%%%%%%%%%%%%%%%%%%%%%%%%%%%%%%%%%%%%%%%%%%%%%%%%%%%%%%%%%%%%
\section{$5d$ diffeomorphism and gauge transformation}
%%%%%%%%%%%%%%%%%%%%%%%%%%%%%%%%%%%%%%%%%%%%%%%%%%%%%%%%%%%%%%%%
In this section we address the issue on the choice of
brane-shift function in a warped metric background. We consider the
perturbed metric in the following 
form\footnote{Here to make a connection to~\cite{KLR} we follow the metric
signature $(+,+,\cdots,-$) and consider a more general case where the $4d$
hypersurface could be a Minkowski, de-Sitter or Anti de-Sitter $3$-brane.}
\begin{equation}
ds^2= \rho(y) (g_{\mu\nu}+h_{\mu\nu})dx^{\mu}dx^{\nu}
-dy^2=g_{ab}\,dx^a\,dx^b\,,\label{genmetric}
\end{equation}
in axial gauge $h_{a 5}=0$. For generality, at first we do not
restrict the form of $\rho(y)$, but will take a proper account of the 
RS-type background solution $\rho(y)= e^{2A(y)},
\,A(y)=(c-|y|)/\ell$, where $c$ is a parameter associated with the geometry
of curved solution and for a flat Minkowski $3$-brane one can set $c=0$. In
terms of the metric fluctuations $h_{ab}$, the full $5$-dimensional
diffeomorphism read
\begin{equation}
\delta h_{ab}=\hat{\td}_a\,\xi_b+\hat{\td}_b\,\xi_a\,.\label{xiab}
\end{equation}
Here hats represent the parameters defined in the five 
space-time dimensions and $\hat{\td}$ is
the $5d$ covariant differential operator. Then the $5d$ killing
vectors are defined by
$\xi^a=(\xi^\mu\, e^{-2 A(y)},\,-\xi^5)$, and $\xi_a=(\xi_\mu,\,\xi_5)$, and
for a non-vanishing brane tension,
the $5d$ killing equations $\hat{\td}_a\,\xi_b+\hat{\td}_b\,\xi_a=0$
can be expressed as 
\begin{equation}
\hat{\td}_\mu\,\xi_\nu+\hat{\td}_\nu\,\xi_\mu=0,\,\,\hat{\td}_\mu\,\xi_5
+\hat{\td}_5\,\xi_\mu=0,
\,\,\hat{\td}_5\,\xi_5=0\,.\label{diffeo}
\end{equation}
The last equation of~(\ref{diffeo}) implies that $\xi_5=\omega(x^\lambda)$,
which can be identified as a $4d$ scalar (or brane shift function in the
warped background), while, the second equation yields
\begin{equation}
\partial_\mu\xi^5(x) = -\partial_y\,\xi_\mu (x^\lambda,y)
+ (\rho^\p/\rho)\, \xi_\mu(x^\lambda,y)\,.\label{ximu5compo}
\end{equation}
Operating by a differential operator $\td_\nu$ on both sides, we get
\begin{equation}
\partial_y\big[\rho^{-1}\,
(\td_\mu \xi_\nu-\td_\nu \xi_\mu)\big]=0\,,
\label{newximu5compo}
\end{equation}
where we have made use of $(\td_\mu \partial_\nu-\td_\nu \partial_\mu)\,
\omega(x^\lambda)=0$. Eq.~(\ref{ximu5compo}) defines general transformations
in terms of arbitrary small functions of $x$, which take the form
\begin{eqnarray}
\xi^5&=&\hat{\xi}\,^5(x)\,,\\
\xi^\mu(x^\lambda,y)&=&- \rho(y)\,G\,\partial^\mu\,\hat{\xi}\,^5(x^\lambda)+
\beta (x^\lambda)\,,\label{gaugetrans}
\end{eqnarray}
where $\beta^\mu(x)$ is an arbitrary dual vector field, which preserves the
remaining gauge invariance, and G is defined as\footnote{To be more precise
we choose zero as the lower limit on the integral.}
\begin{equation}
G=\int^{\tilde{y}}\,\rho(y)^{-1}\,d\tilde{y}\,.
\end{equation}
Expanding the first equation of~(\ref{diffeo})
\begin{equation}
(\td_\mu \xi_\nu + \td_\nu \xi_\mu)-g_{\mu\nu} \rho^\p \omega(x^\lambda)=0\,,
\label{ximunueqn}
\end{equation}
multiplying this by $\rho^{-1}$, differentiating w.r.t. $y$ and combining
the result with~(\ref{newximu5compo}), we find
\begin{equation}
\omega^{-1}\,\td^2 \omega=-2 \rho\,\big(\rho^\p/\rho\big)^\p\,.\label{scalar}
\end{equation}
For the RS background solution $\rho=e^{-2|y|/\ell}$, the r.h.s. is
$8\,\ell^{-1}\,e^{2|y|/\ell}\,\delta(y)$, which is non
vanishing at $y=0$. Further, since the l.h.s. of Eq.~(\ref{scalar}) is a 
function of $x^\lambda$ and the r.h.s. is a function of just $y$, by 
separation of variables, one requires
\begin{equation}
(\td^2+k)\,\omega=0\,, ~~~ k-2\,\rho\,(\rho^\p/\rho)^\p=0\,.
\label{omegaeqn}
\end{equation}
Here $k= -8\,\ell^{-1}\,e^{2|y|/\ell}\,\delta(y)$, so that $k$ vanishes
in the bulk. Since $k(y=0)\neq 0$ and $k(y\neq 0)=0$, it may be 
inconsistent in requiring a vanishing $w$ at the brane, but a non-vanishing
$\omega$ in the bulk. So, in general, we require a non-vanishing $\omega$
both on the brane and in the bulk.

Identifying $\hat{\xi}\,^5(x)=-\Phi$ as the radion field~\cite{KLR}
associated to the fifth coordinate transformation and
using $\rho(y)=e^{2A(y)}$, we get
\begin{equation}
\td^2 \Phi=4\lambda \Phi\,,\label{waveeqn}
\end{equation}
where $\lambda =- e^{2A(y)}\,A^{\p\p}$ is defined as the $4d$
cosmological constant and $R_{(4)}=- 4\lambda$. This can
be attributed to the the weak energy condition ($i.e., T_0^0-T_5^5\geq 0$):
$A^{\p\p}\leq -\lambda e^{-2A}$ obtainable from the Einstein field
equations, and this was first invoked in~\cite{KLR}. 
The stronger version $A^{\p\p}<0$ of the standard c-theorem then requires
$\lambda >0$ and hence implies an $AdS_4$ brane ($R_{(4)}<0)$ embedded in
$AdS_5$ bulk (i.e., in a constant negative curvature background). This leads
to locally localized gravity for a sufficiently small
$AdS_4$ cosmological constant studied in~\cite{KLR} (see~\cite{Kogan2} for a 
bigravity model with two positive tension $AdS_4$ branes in $AdS_5$), 
though for a more meaningful four-dimensional physics we might expect a 
completely localized gravity on the brane, if the spin-$2$ graviton is 
strictly massless. A flat Minkowski $3$-brane with a singular
($\delta$-function) source is, however, compatible with the standard 
$c$-theorem, viz. for a vanishing $4d$ cosmological constant ($\lambda=0$), 
one acquires $A^{\p\p}=-2\,\ell^{-1}\delta(y)\leq 0$.

Finally, substituting Eq.~(\ref{ximu5compo}) into Eq.~(\ref{ximunueqn}), 
we find 
\begin{equation}
\td_\mu\,\beta_\nu+\td_\nu\,\beta_\mu=g_{\mu\nu}\,(\rho^\p/\rho)\,\omega
+2\, G\, \td_\mu\,\partial_\nu \omega\,.\label{ximu5plusmunu}
\end{equation}
Contracting the space-time indices and using
$(\td^2+k)\omega=0$, this gives
\begin{equation}
\omega^{-1}\,\td_\mu\,\beta^\mu=2 \big[(\rho^\p/\rho)-k\big]\,.\label{betaeqn}
\end{equation}
Since $\beta^\mu$ is the killing vector associated with four dimensional
space-times, by separation of variables, one requires
$(\rho^\p/\rho)-k=constant$.
This demands that the warp factor $\rho(y)$ takes a form
$\rho(y)=\rho(0)\,e^{k_1 y}$. This is also justified from the second equation
of~(\ref{omegaeqn}). At the location of the brane we
choose $\rho(0)=1$ and define $k_1=-2/\ell$ to arrive at
$\rho(y)=e^{-2|y|/\ell}$. This implies that one may set
$\omega(y=0)=0$ only for $\rho(y)\neq \rho(0)\,e^{2|y|/\ell}$, but not for
the RS-type background solution. In particular, with the choice $\omega(y=0)=0$
one cannot solve the killing equations without breaking the $D$-dimensional
diffeomorphism. With $\omega=0$, one gets the $4d$ killing equations 
$\pounds_\beta g_{\mu\nu}=\nabla_\mu\beta_\nu+\nabla_\nu\beta_\mu=0$. 

In the RS set-up, as the coordinate systems are based on the $4d$
hypersurface, in the presence of matter fields localized on the brane, the
brane would shift to a new location or appear bent. There will, however,
not be a global coordinate system which is Gaussian-normal in the latter
case. Therefore, a brane bending mechanism, which may be important to study the
gravitational interactions between matter sources localized on the brane, is
needed to preserve the axial gauge, $h_{a5}=0$. The latter is also
useful to remove the gauge degrees of freedom completely. Though the
``brane-bending'' mechanism may not be so obvious when one works with
different gauge~\cite{IYA, KAKU}, we find still more convenient to work
with Gaussian-normal conditions.

Let's consider the ground state metric in a more relevance form
\begin{equation}
ds^2=dy^2+e^{-2|y|/\ell}\,(g_{\mu\nu}+h_{\mu\nu})\,dx^\mu\,dx^\nu\,.
\label{RSbgmetric}
\end{equation}
When one deforms the coordinates changing the base hypersurface to a new
set of coordinates $(x', y')$ as
\begin{equation}
y'=y+\hat{\xi}\,^5(x)\,,~~
x^{\mu'}=x^\mu+
\xi^\mu(x^\lambda,y)=x^\mu-\frac{\ell}{2}\,e^{2|y|/\ell}\,
\partial^\mu\,\hat{\xi}\,^5(x)+
\beta^\mu (x^\lambda)\,,\label{coor.trans}
\end{equation}
the graviton fluctuations $h_{\mu\nu}$, under these coordinate
transformations, transform as
\begin{equation}
h_{\mu\nu}(x)\to  h^{\p}_{\mu\nu}(x)
=h_{\mu\nu}(x)+\td_\mu\xi_\nu+\td_\nu\xi_\mu
-2\,g_{\mu\nu}\,\ell^{-1}\hat{\xi}\,^{5}(x)\,.
\label{hmunusum}
\end{equation}
For a flat Minkowski $3$-brane, $g_{\mu\nu}= \eta_{\mu\nu}$,
the above transformation for the metric fluctuations reads 
\begin{equation}
h_{\mu\nu}(x)\to  h^{\p}_{\mu\nu}(x)
=h_{\mu\nu}(x)-\ell^{-1}\,\partial_\mu\partial_\nu\,\hat{\xi}\,^5(x^\lambda)
+e^{-2|y|/\ell}\big(\partial_\mu\beta_\nu+\partial_\nu\beta_\mu
-2\,\eta_{\mu\nu}\,\ell^{-1}\hat{\xi}\,^{5}(x)\big)\,,
\label{hmunutrans}
\end{equation}
where the first and second terms in the bracket are pure gauge terms in the 
$4$-dimensional brane worldsheet, and can be gauged away by defining
$\partial_\mu\partial_\nu\,\hat{\xi}\,^5(x)$ appropriately.
%%%%%%%%%%%%%%%%%%%%%%%%%%%%%%%%%%%%%%%%%%%%%%%%%%%%%%%%%%%%%%%%%%%%%%%%
\section{Linearized equations and graviton propagators}
For the action~(\ref{action}), the linearized equations of motion read as
\begin{equation}
\delta\hat{G}_{ab}+\kappa\, \delta^{(1)}\hat{H}_{ab}
=\frac{\kappa}{2} T_{ab}^{(m)}\,,\label{hatequation}
\end{equation}
where
$\delta \hat{G}_{ab}=\delta G_{ab}-\delta {T^{(0)}}_{ac}\, h_b\,^c$ and
$\delta \hat{H}_{ab}= \delta H_{ab}- H_{ac}\, h_b\,^c$.
The metric~(\ref{conmetric}) would be more convenient to simplify the
analysis of gravitational fluctuations. Thus
we consider the metric perturbations in the form
$g_{ab}= e^{-2A(z)}\big(\eta_{ab}+h_{ab}\big)$. In axial gauge,
$h_{a5}=0$, to the first order in $h_{ab}$, the tensor mode of the
metric fluctuation $h_{\mu\nu}$ satisfies~\cite{IPN1}
%%%%%%%%%%%%%%%%%%%%%%%%%%%%%%%%%%%%%%%%%%%%%%%%%%%%%%%%%%%%%%
\begin{eqnarray}
&&\left[-\big(1+4\alpha\kappa e^{2A} A^{\p\p}\big)\partial_\lambda^2
-\big(1-4\alpha\kappa e^{2A} {A^\p}\,^2\big)\,\partial_z^2
+3\,A^\p  \partial_z\right]\big(h_{\mu\nu}-\eta_{\mu\nu} h\big)\nn\\
&&+\left(1-4\alpha\kappa e^{2A}\left(2{A^\p}\,^2-A^{\p\p}\right)\right)
\left(2\partial_{(\mu}\partial^\lambda h_{\nu)\lambda}
-\partial_\mu \partial_\nu h \right) 
- \left(1+4\alpha\kappa e^{2A} A^{\p\p}\right)\times\nn\\
&&\eta_{\mu\nu}\partial_\lambda\partial_\rho
h^{\lambda\rho} +4\alpha\kappa e^{2A} A^\p
\left[\left(2A^{\p\p}-{A^\p}\,^2\right) \partial_zh_{\mu\nu}
+\left(A^{\p\p}-2{A^\p}\,^2\right)\eta_{\mu\nu} \partial_zh\right]
=\kappa T^{(m)}_{\mu\nu}\label{withttcomp}\,.
\end{eqnarray}
%%%%%%%%%%%%%%%%%%%%%%%%%%%%%%%%%%%%%%%%%%%%%%%%%%%%%%%%%%%%%%%%%%%%%%%%%%
One can look for solutions of the form
$h_{\mu\nu}(x,z)=\epsilon_{\mu\nu}\,
e^{ipx}\,\psi(z)$. Here $\epsilon_{\mu\nu}$ is the constant polarization
tensor of the graviton wave function, $m^2=-p^2$ and $m(=\sqrt{-p\cdot p})$ is
the four-dimensional mass of the perturbation. 
Consider first the case without matter source on the brane. Then with the
RS background solution $A=\log (|z|/\ell+1)$, the TT components of 
the metric fluctuations would imply the following expression of graviton 
propagator
\begin{eqnarray}
&&\bigg[\left(1-\frac{4\alpha\kappa}{\ell^2}\,sgn(z)^2
+\frac{8\alpha\kappa}{\ell}\delta(z)\right)\,
\partial_\lambda^2+\left(1-\frac{4\alpha\kappa}
{\ell^2}\,sgn(z)^2\right)\,\partial_z^2 -\frac{16\alpha\kappa}{\ell^2}
sgn(z)\,\delta(z)\, \partial_z\nn\\
&&~ -\frac{3}{(\ell+|z|)}
\left(1-\frac{4\alpha\kappa}
{\ell^2}\,sgn(z)^2\right)\, sgn(z)\,\partial_z
\bigg]\, \G_5(x,z;x',z')
=\delta^{(4)}(x-x')\delta(z-z')\,.
\label{greenpro1}
\end{eqnarray}
The graviton propagator along the brane can be decomposed into the Fourier
modes
\begin{eqnarray}
\G_5(x,z;x^\p, z^\p)=\int\frac{d^4 p}{(2\pi)^4}
e^{ip(x-x^\prime)}\G_p(z,z^\p),\label{fourier}
\end{eqnarray}
where the Fourier components $\G_p(z,z')$ in the bulk satisfy
\begin{equation}
e^{2A}\big(\partial_z^2 - p^2-\frac{3}{z}\partial_z\big)\G_p(z,z^\prime)
=e^{5A}\big(1-\gamma\big)^{-1} \delta(z-z^\prime)
\label{propa1}\,.
\end{equation}
Since $e^{A(z)}=(|z|/\ell+1)$, one
can use $e^{A(z)}=z/\ell$ for $|z|>>\ell$.
Writing $\G_p = \big(z z'/ \ell^2\big)^2\, \hat{\G}_p$ and
$p^2 = -q^2$, we arrive at
\begin{eqnarray}
\big(z^2\partial^2_z+z\partial_z+q^2 z^2-4\big)\hat{\G}_p(z,z^\p)
=\big(1-\gamma\big)^{-1} \ell\, z\,\delta(z-z^\p)
\label{greeneq2}\,.
\end{eqnarray}
The translation invariance of $\hat{\G}_p(z, z')$ implies that this Bessel
equation is generally valid after a general coordinate transformation in
Gaussian-normal form. When one defines a coordinate
transformation $z\,=\,\ell\,e^{y/\ell}$ and works in the RS background,
for $z>\ell$ (or $z<\ell$) the metric is the $5$-dimensional $AdS$ metric
given by
\begin{equation}
dS^2=\frac{\ell^2}{z^2}(dx_4^2 +dz^2)\,.\label{rsmetric}
\end{equation}
So without loss of generality, one can locate the brane at $z=\ell$, and
the brane at $y=0$ is mapped to $z=\ell$. Therefore, in $y$-coordinate,
the TT components of the metric fluctuations
satisfy\footnote{ See also ref.\cite{KL}.}
\begin{eqnarray}
&&\Big[\big(1-4\alpha\kappa{A^\p}\,^2+4\alpha\kappa A^{\p\p}\big)
e^{2|y|/\ell}\,\partial_\lambda^2
+\partial_{y}\,^2 - 4\alpha\kappa\,\big({A^\p}\,^2\,
\partial_{y}\,^2+ 2A^\p\,A^{\p\p}\,\partial_{y}\big) ~~~~~~~~~~~\nn\\
&& ~~-4{A^\p}\,^2\big(1-4\alpha\kappa {A^\p}\,^2\big)
+2 A^{\p\p}\big(1-12\alpha\kappa {A^\p}\,^2\big)\Big]\,h_{\mu\nu}(x,y)
=-\kappa\,T_{\mu\nu}(x,y)
\label{linear-in-y}\,.
\end{eqnarray}
Here the energy momentum tensor includes a contribution from matter
source on
the brane, $T_{\mu\nu}(x, y)= S_{\mu\nu}(x)\,\delta(y)$. Now we deform
the coordinates from $(x, y)$ to $(x', y')$. In the latter gauge the brane
is located at
$y'=y+\xi^5(x)=0$, where $\xi^5(x)$ is an arbitrary brane shift function
defined in the Section 3. In $y'$-coordinate, one has
\begin{eqnarray}
&&\Big[\big(1-4\alpha\kappa{A^\p}\,^2+4\alpha\kappa A^{\p\p}\big)
e^{2|y|/\ell}\,\partial_\lambda^2
+\partial_{y}^2 - 4\alpha\kappa\,\big({A^\p}\,^2\,
\partial_{y}^2+ 2A^\p\,A^{\p\p}\,\partial_{y}\big) ~~~~~~~~~~~\nn\\
&& ~~-4{A^\p}\,^2\big(1-4\alpha\kappa {A^\p}\,^2\big)
+2 A^{\p\p}\big(1-12\alpha\kappa {A^\p}\,^2\big)\Big]\,h^\p_{\mu\nu}
=-\kappa\Sigma_{\mu\nu}(x')\delta(y')
\label{TTlinear-in-y'}\,,
\end{eqnarray}
where the source term $\Sigma_{\mu\nu}(x')$ is given by (see Appendix A)
\begin{equation}
\Sigma_{\mu\nu}(x')=S_{\mu\nu}(x')-\frac{1}{3}\Big(
\eta_{\mu\nu}-\frac{\partial_\mu\, \partial_\nu}
{\partial_\lambda^2}\Big) S(x')\,.
\label{sourceterm}
\end{equation}
The condition $\Sigma_\mu^\mu=0$ justifies the gauge
choice $h'=0$, and a factor of $(1+\gamma)$ in the second and third terms due
to the Israel junction condition for $h^\p_{\mu\nu}$ has been canceled by the
term $(1+\gamma)^{-1}$ coming from the brane-shift function.

Now for the TT components of the metric fluctuations, the Fourier modes 
$\hat{\G}_p$ of the graviton propagator satisfy
%%%%%%%%%%%%%%%%%%%%%%%%%%%%%%%%%%%%%%%%%%
\begin{eqnarray}
&&\left(1-\frac{4\alpha\kappa}{\ell^2}sgn(y)^2
+\frac{8\alpha\kappa}{\ell}\delta(y)\right)e^{2|y|/\ell}\,
\partial_{\lambda}^2{\hat \G}_p+\partial_y^2\,\hat{\G}_p
- \frac{4\alpha\kappa}{\ell^2}\,
\left(sgn(y)^2\,\partial_y^2+2\delta(y) sgn(y)\partial_y\right)
\hat{\G_p}\nn\\
&&-\frac{4}{\ell^2}sgn(y)^2\,\bigg(1-\frac{4\alpha\kappa}{\ell^2}
sgn(y)^2\bigg)
{\hat \G}_p+\frac{4}{\ell}\delta(y)\,\big(1-\frac{12\alpha\kappa}{\ell^2}\,
sgn(y)^2\big){\hat \G}_p = \delta(y-y')
\label{fourieriny}\,.
\end{eqnarray}
Indeed, a choice of appropriate boundary conditions (b.c.)
on the brane is crucial in obtaining the correct behavior of the graviton
propagators. With the GB term, in particular, there exists a subtlety in the
choice of boundary condition
for $\hat{\G}_p$ across the brane due to the terms involving
$\partial_y|y|$ at
$y=0$. But one should fix them from the requirements that one obtains a
consistent low-energy limit and the theory becomes free from any unphysical
(ghost) states. For this reason, one has to properly regularize the terms 
such as $sgn(y)^2\,\partial_y^2\hat{\G}$, $sgn(y)^2\,\delta(y)$ and 
$sgn(y)\delta(y)\,\partial_y\hat{\G}$. 
It should be understood that $\partial_y \G_p (|0|)=0$, but 
$\partial_y\G|_{y=0_+}=-\partial_y\G|_{y=0_-}\neq 0$. And, just below or
above $y=0,\,\,\delta(y)$ still dominates $sgn(y)$, the latter shows a
behavior of a step function. In other words, the function $sgn(y)$ is defined 
to vanish for vanishing argument and only for $y\neq 0$ one can use 
$sgn(y)^2=1$, thus $sgn(y)=+1$ if
$y>0$ and $sgn(y)=-1$ if $y<0$. The term $sgn(y)\,\partial_y \hat{\G}_p$ 
vanishes when evaluated from just below to just above the brane.  We also 
need the following two equalities, obtained after properly regularizing the 
$\delta$-function, to simplify the above equation   
$$
\delta(y)\,sgn(y)^2 =\delta(y)/3\,,$$
$$
\left(sgn(y)^2\,\partial_y^2+2\delta(y) sgn(y)\partial_y\right)\hat{\G_p}=
\partial_y^2 \hat{\G_p}(|y|)+2\,\delta(y)\partial_y \hat{\G_p}(0_+)\,.
$$
Given these 
considerations, the boundary condition at $z=\ell$ is given by\footnote{ 
See also Ref.~\cite{IPN4} for a rigorous derivation of this boundary 
condition.} 
\begin{equation}
\bigg(z\partial_z +2+\chi\, 
q^2 z^2\bigg)\hat{\G}_p(z,z^\p)|_{z=\ell}=0\,,
\label{greenbc}
\end{equation}
where $\chi=\gamma/(1-\gamma)$. Eq.~(\ref{greeneq2}) also implies the following matching conditions at
$z=z'$:
\begin{eqnarray}
\hat{\G}_<|_{z=z^\p}&=&\hat{\G}_>|_{z=z^\p},\nn\\
\partial_z(\hat{\G}_>-\hat{\G}_<)|_{z=z^\p}&=& \big(1-\gamma\big)^{-1}
\frac{\ell}{z^\p}
\label{match1}\,.
\end{eqnarray}
The general solutions of the Bessel equation~(\ref{greeneq2}) satisfying 
the b.c.~(\ref{greenbc}) and the matching Eqs.~(\ref{match1}) lead to the
following expression for the graviton propagator, for $z < z'$,
\begin{equation}
\hat{\G}_{z<z'} = iA(z^\prime)\bigg[\bigg(J_1(q\ell)+
\chi\,q\ell J_2(q\ell)\bigg)
H^{(1)}_2(qz)-\bigg(H^{(1)}_1(q\ell)+\chi\,q\ell H^{(1)}_2(q\ell)
\bigg)J_2(qz)\bigg]
\label{greenleft}\,,
\end{equation}
where $H^{(1)}_{1,2}=J_{1,2}+iY_{1,2}$ is the Hankel function of the first
kind. For $z>z'$ the use of the b.c. as $z\to\infty$, similar to the
Hartle-Hawking b.c. which requires the $+ve$ frequency wave be ingoing to
the $AdS$ horizon $z\to\infty$ but no re-emission, results in
\begin{equation}
{\hat \G}_{z>z'}\,=\,B(z')\,H^{(1)}_2(qz)
\label{Hawkingbc}\,.
\end{equation}
The use of Eq.~(\ref{match1}) would imply the following general
expression for the graviton propagator~\cite{GKR}:
%%%%%%%%%%%%%%%%%%%%%%%%%%%%%%%%%%%%%%%%%%%%%%%%%%%%%%%%%%%%%%%%%%%%%%
\begin{eqnarray}
\G_5(x,z;x,z^\p)&=&(1-\gamma)^{-1}\frac{i\pi}{2\ell^3}(z z^\p)^2
\int\frac{d^4p}{(2\pi)^4} e^{ip(x-x')}\nn\\
&\times&\left[\left(\frac{J_1(q\ell)
+\chi\,q\ell J_2(q\ell)}{H^{(1)}_1(q\ell)+\chi\,q\ell H^{(1)}_2(q\ell)}
\right) H^{(1)}_2(qz) H^{(1)}_2(qz') - J_2(qz_<)H^{(1)}_2(qz_>)\right]
\label{greenpro2}\,.
\end{eqnarray}
The second term implies a presence of extra polarization state as
massless $4d$ scalar field in the full $5d$ graviton propagator. One 
may inquire whether this contributes to the propagator on the brane.
This issue was discussed in \cite{CEH, DGP, RGV}, and it has been
argued in \cite{CEH} that this extra polarization degree of freedom in the
full propagator may be cancelled by the brane-bending mode
\footnote{However, the brane-bending effect in order to cancel this
unwanted physical polarization of gravitons due to the mismatch in the
tensor structure of massive and massless graviton propagators is not free
of comment \cite{RGV}.}, and this is indeed precisely correct in an 
appropriate gauge (see below) even in the presence of the 
GB interaction term.

When one of the arguments of $\G_5$ is at $z'=\ell$,
the above graviton propagator reduces to
\begin{equation}
\G_5(x,z;x',\ell)=(1-\gamma)^{-1}\frac{z^2}{\ell^2}
\int\frac{d^4 p}{(2\pi)^4} e^{ip(x-x')}\frac{1}{q}
\left[\frac{H^{(1)}_2(qz)}{H^{(1)}_1(q\ell)
+\chi\, q\ell H^{(1)}_2(q\ell)}\right]
\label{greengen}\,.
\end{equation}
For both points at $z, z'= \ell $, use of the recursion relation
$H^{(1)}_0(q\ell)+H^{(1)}_2(q\ell)= (2/q\ell) H_1^{(1)}(q\ell)$ would imply
%%%%%%%%%%%%%%%%%%%%%%%%%%%%%%%%%%%%%%%%%%%%%%%%%%%%%%%%%%%%%%%
\begin{equation}
\G_5(x,\ell;x', \ell)={(1-\gamma )}^{-1}
\left[\frac{2}{\ell}\, \Delta_4(x,x^\prime)
+\Delta_{KK}(x,x')\right]
\label{greensum},
\end{equation}
where
\begin{eqnarray}
\Delta_4(x,x^\prime)&=&\int\frac{d^4 p}{(2\pi)^4} \frac{e^{ip(x-x^\prime)}}
{q^2}\,\left[\frac{H_1^{(1)}(q\ell)}
{H_1^{(1)}(q\ell)+\chi\,q\ell\,H_2^{(1)}(q\ell)}\right]\nn\\
&\approx&(1+2\chi)^{-1} \int\frac{d^4 p}{(2\pi)^4}\,e^{ip(x-x^\prime)}\,
\left[\frac{1}{q^2}-\frac{\chi}
{1+2\chi}\,\ell^2\,\left(\ln(q\ell/2)+\Gamma\right)
\right]+\cdots,\label{green4d}
\end{eqnarray}
\begin{eqnarray}
\Delta_{KK}(x,x^\prime) &=& -\int\frac{d^4 p}{(2\pi)^4}
e^{ip(x-x^\prime)}\times \frac{1}{q}
\left[\frac{H^{(1)}_0(q\ell)}{H^{(1)}_1(q \ell)
+\chi\,q\ell\, H^{(1)}_2(q\ell)}\right]\nn\\
&\approx& (1+2\chi)^{-1}\int\frac{d^4 p}{(2\pi)^4}\,
e^{ip(x-x^\prime)}\,\ell\,\left[\ln(q\ell/2)+\Gamma\right]+
{\cal O}\big[(q\ell)^3\big]\,,\label{kkgreen}
\end{eqnarray}
where $\Gamma=0.577216\cdots$ is the Euler-Mascheroni constant. Clearly, for 
$\chi=0$, there is no sub-leading term from $\Delta_4(x,x')$
and the correction term arises only from $\Delta_{KK}(x,x')$. However, for
a non-vanishing and relatively large $\chi\,(\lesssim 1)$, the sub-leading
term from $\Delta_4$ could be in the order of leading order term from
$\Delta_{KK}$. In order to
expect a dominant contribution from
$\Delta_{KK}(x,x')$ as a correction term of the Newtonian potential, one 
may require a small $\gamma$. This indeed imposes a large mass hierarchy limit
$\ell M_* >> 1$.

Substituting Eqs.~(\ref{green4d}, \ref{kkgreen}) into Eq.~(\ref{greensum}) or
using the expansion:
\begin{equation}
\frac{H_2(q\ell)}{H_1(q\ell)+\chi\,q\ell\,H_2(q\ell)}
\sim\frac{1}{1+2\chi}\,\left(\frac{q\ell}{2}+\frac{2}{q\ell}\right)+
\frac{q\ell}{(1+2\chi)^2}
\,\left[\ln(q\ell/2)-\chi+(\Gamma-1/2)\right]+{\cal O}\left((q\ell)^3\right)\,,
\label{expansion}
\end{equation}
for $|x-x'|>>\ell$, i.e., $q\ell<<1$, the propagator~(\ref{greensum})
is given by, to the leading order terms,
\begin{eqnarray}
\G_5(x,z;x',\ell)&\approx&(1-\gamma)^{-1}(1+2\chi)^{-1}
\int\frac{d^4 p}{(2\pi)^4}\, e^{ip(x-x')}\,\left[\frac{2}{\ell}\,\frac{1}{q^2}
+\frac{1}{1+2\chi}\,\ell \ln(q\ell/2)\right]\nn\\
&=&(1+\gamma)^{-1}
\left[\frac{2}{\ell}\, \G_4(x,x^\prime)
+\G_{KK}(x,x')\right]\,,\label{fullgreen}
\end{eqnarray}
where $\chi=\gamma/(1-\gamma)$ is substituted. In physical sense, the 
correction term $(1+2\chi)^{-1}$ arises from the 
coupling between GB curvature term and the brane matter. Evidently, the 
zero-mode graviton propagator on the brane is given by 
\begin{equation}
\G_4(x,x')=\int\frac{d^4p}{(2\pi)^4}\,\frac{1}{q^2}
= \square_4(x,x')\label{g4delta4}\,,
\end{equation}
while
$\square_4(x,x')$ is the usual $4d$ flat space (massless) scalar propagator.
Eq.~(\ref{greensum}), therefore, suggests that even with the
GB interaction term, in the low-energy scale $q\ell<<1$, the zero-mode
propagator reproduces a correct $4d$ massless graviton propagator. Also the 
zero-mode graviton propagator on the brane satisfies
\begin{equation}
\partial_\mu\partial^\mu\,\G_4 (x, x')= \delta^4 (x, x')\,.
\label{truezeromode}
\end{equation}
Obviously, the overall constant term $(1+\gamma)^{-1}$ renormalizes the $4d$ 
Newton's constant on the brane. The limiting behavior of the graviton 
propagator is given by
\begin{eqnarray}
\G_4(x,x^\prime)&=&\int\frac{d^4 p}{(2\pi)^4}
\frac{e^{ip(x-x^\p)}}{q^2}
\propto \frac{1}{|x-x'|^2}\,,\label{green4dnew}\\
%\end{eqnarray}
%\begin{eqnarray}
\G_{KK}(x,x')&=&(1+2\chi)^{-1}\,\ell\,
\int\frac{d^4 p}{(2\pi)^4}\,
e^{ip(x-x^\p)}\, \ln(i p\ell/2)
\propto \frac{(1-\gamma)}{(1+\gamma)}\,
\frac{\ell}{|x-x'|^4}\,.\label{gkknew}
\end{eqnarray}
Clearly, at large distances along the brane $|x-x'|=r>>\ell$,
contribution from the KK kernel~(\ref{gkknew}) would be
very small compared to the zero mode piece~(\ref{green4dnew}), and one also 
finds a smooth infra-red behavior of the propagators.

The very short distance $r<<\ell$ behavior of the propagator on the
brane is governed
by the ultra large $q$-behavior of the Fourier mode. In this case
$q\ell>>1$, then using the asymptotic expansion of the Hankel functions, 
one finds from Eq.~(\ref{greensum}), to the leading order,
\begin{equation}
\Delta_4(x,x')\approx -(1+2\chi)^{-1}\int\frac{d^4p}{(2\pi)^4}\,
e^{ip(x-x')}\,\frac{\Delta m}{p^2(p+\Delta m)}\,,\label{4dnewhigh}
\end{equation}
\begin{equation}
\Delta_{KK}(x,x')\approx (1+2\chi)^{-1}\int\frac{d^4p}{(2\pi)^4}\,
e^{ip(x-x')}\,\frac{\Delta m}{p(p+\Delta m)}\label{4dkkhigh}\,,
\end{equation}
where
\begin{equation}
\Delta m\,\equiv\,\frac{1}{\ell}\bigg[2+\frac{\ell^2 M_*^2}{4\alpha'}
\bigg],\,\,\,\,\alpha'\neq 0\label{delta-m}\,.
\end{equation}
Even in the ultra-violet range, the graviton propagators do not blow up.

%%%%%%%%%%%%%%%%%%%%%%%%%%%%%%%%%%%%%%%%%%%%%%%%%%%%%%%%%%%%%%%%
\section{(In) stability of the linearized solutions}
In fact, though the propagator is a good first
test of the linearized approximation, it sometimes does not contain all
information about the graviton spectrum. So it may be the case that one does
not notice an instability in the propagator on the brane, but there are still
unstable modes in the spectrum. In order to judge this we would like to
see whether there is any unphysical (ghost) states in the solutions at the
linearized level.
%%%%%%%%%%%%%%%%%%%%%%%%%%%%%%%%%%%%%%%%%%%%%%%%%%%%%%%%%%%%%%%%%%%%%%%%%%
\subsection{Extra polarization of GB term}
Let us briefly review the results in Ref.~\cite{KL}~\footnote{Addendum: 
however, an importance of $\delta$-function regularization and some 
needed corrections on those results were later narrated by the authors 
of~\cite{KL} in Erratum-ibid, Nucl. Phys. B {\bf 619} (2001), 763.}, 
where metric fluctuation on the brane at 
$z=\ell$ was evaluated with the result (one has to replace $\beta$
in~\cite{KL} by $\gamma/(1-3\gamma)$ in our notation)
\begin{eqnarray}
h_{\mu\nu}(x)&=&-M^{-2}_{pl}\int d^4 x^\p\,\square_4(x,x^\p)
\bigg[S_{\mu\nu}(x^\p)-\bigg(\frac{1}{2}+\frac{\gamma}{3(1-3\gamma)}\bigg)
\eta_{\mu\nu}\, S^\lambda_\lambda(x^\p)\bigg]\nn\\
&&-\frac{M_*^{-3}}{2}\,\frac{(1-\gamma)^{-1}}{(1-3\gamma)}
\int d^4 x^\prime\, \G_{KK}(x,x^\p)
\bigg[S_{\mu\nu}(x^\p)-\frac{1}{3} \eta_{\mu\nu}\,
S^\lambda_\lambda(x^\prime)\bigg]\,.
\label{totalfluc}
\end{eqnarray}
In Eq.~(\ref{totalfluc}), in particular, 
$M_{pl}^2= M_*^3\,\ell\,(1-\gamma)^{-2}(1-3\gamma)$
is to be understood. For $\gamma=0$,
contribution from the brane-bending mode changes the factor $1/3$ to
$1/2$ in Eq.(\ref{totalfluc}), and hence yields the usual
massless $4d$ graviton propagator
\begin{equation}
\square_4(x,x')_{\mu\lambda\nu\rho}=\int\frac{d^4p}{(2\pi)^4}
\frac{e^{ip\cdot(x-x')}}{2 p^2}\big(g_{\mu\nu}g_{\lambda\rho}+
 g_{\mu\rho}g_{\lambda\nu}- g_{\mu\lambda}g_{\nu\rho}\big)\,.
\label{graviprop}
\end{equation}
However, for a non-trivial $\gamma$, the usual $4d$ gravity appears
to be modified by the extra polarization factor $-\gamma/3(1-3\gamma)$, and
the coupling $\gamma$ here induces an extra polarization/ghost 
state that cannot
be resolved by adjusting any physical parameters. For the
result~(\ref{totalfluc}), the leading behaviors of
$h_{00}$ and $h_{ij}$ components will also be different for a
non-vanishing $\gamma$. In this case, one essentially encounters massive or
massless ghosts, which delocalize gravity at the background. This implies
that the zero mode of the quantum mechanical system becomes unstable due to
the higher-curvature terms even in the GB combination. In other words, the 
extra polarization state in the graviton propagator reveals that one
does not recover a correct tensor structure while admitting the brane brane
bending
effect in the presence of the GB term. As the brane bending effect
itself results from the presence of a localized source on the brane, this
might signal out a conflict of the localized source in the
presence of the GB interaction.

One of the main points of this paper is to argue that none of the statements
about the ghost/extra polarization state in the previous paragraph is 
necessarily true. As we know, the GB term is a ghost free combination 
independent of the dimensions and the topology of the space-time, and 
there is no ghost state in the RS model 
when $\alpha=0$~\cite{GTT}. Thus we 
do expect to recover a correct momentum behavior (i.e., $4d$ Newton's 
law), and also a correct tensor structure (i.e., graviton polarization 
state) in the RS single brane model. This is actually what we find 
by correctly defining the Neumann boundary condition 
(Israel junction condition, and enforcing $Z_2$ symmetry) on the brane, 
which requires a proper regularization of the $\delta$-function. 
In doing so the extra polarization state simply does not arise, and 
hence and one find both--a correct momentum behavior and  also a 
correct tensor structure for the massless graviton propagators on the 
brane. 

At any rate, to visualize the negative norm (for $\gamma>0$) state for the
result~(\ref{totalfluc}), one can evaluate the one-particle exchange 
amplitude in the presence of matter source, $T_{\mu\nu}^{(m)}$. This is 
given by
\begin{equation}
{\cal A} =\frac{1}{2}\int_{\partial M}\sqrt{-g}\,
h_{\mu\nu}(x) T^{\mu\nu}_{brane}(x)
\equiv\frac{8\pi G_N}{p^2}\bigg[S_{\mu\nu}S^{\mu\nu}
-\bigg(\half+\frac{\gamma}{3(1-3\gamma)}\bigg) S^2\bigg]\,,
\label{tbrane}
\end{equation}
which, in the massless limit, yields
\begin{eqnarray}
{\cal A}_{massless}=\frac{8\pi\bar{G}_N}{-p^2_0+p^2_3}
\bigg[|S_{+2}|^2+|S_{-2}|^2
-\frac{\gamma}{3(1-3\gamma)}(S_{11}+S_{22})^2\bigg]\,.\label{mtendtozero}
\end{eqnarray}
The first two terms, $S_{\pm2}=\half(S_{11}-S_{22})$, are the
contributions from the gravitons with helicities
$\lambda=\pm 2$, which represent the massless spin-2 propagation, and the
last term is due to the massless graviscalar. So one needs either
$S_{11}+S_{22}=0$ or $\gamma<0$ for no negative norm state. The requirement
$S_{33}=2(S_{11}+S_{22})=0$ is invalid, because this implies 
$S_\mu^\mu=0$, but $S_{11}+S_{22}\neq 0$ in general. Further, by 
demanding a negative $\gamma$, if one requires definite-positive 
norms, one will introduce an unphysical state into the theory due to
the presence of a graviscalar mode and this further implies that the gauge
$h_{55}=0$ could be an insufficient gauge for EGB gravity. In particular,
this may suggest that the cancellation of an unphysical scalar mode by the
brane bending effect is incomplete in the presence of higher-curvature
terms. But, this is indeed not the case as we see below.

It may perhaps be argued that the above residual effect of the 
graviscalar mode may have been due to the ignorance of the $T_{55}$ 
component or a choice of Gaussian normal condition 
$h_{55}=h_{\mu 5}=0$, the former (a non-trivial $T_{55}$) can 
arise from the
physics responsible for the stabilization of fifth space (as implicitly
implied by the energy conservation in the bulk), and the latter is used to
completely remove the gauge degrees of freedom in the RS set-up. 
However, even a non-trivial contribution of the $T_{55}$ component, which
generally need not vanish at the quantum level, does not 
appear to remedy the ghost (negative norm) state of the above nature. 
In fact, it can be easily seen that an apparent presence of ghost state in 
the above observations was simply due to an incorrect boundary condition 
imposed on the brane at $y=0$, and for properly regularized $\delta$ 
function, the theory is completely free from ghost (negative norm) state. 
This is the case we discuss below.
%%%%%%%%%%%%%%%%%%%%%%%%%%%%%%%%%%%%%%%%%%%%%%%%%%%%%%%%%%%%%%%%%%%%%%%%%%%
\subsection{No ghost-state of GB term}
By taking into proper accounts of the non TT components, we find that a net
result that growing part of $h$ can be eliminated by a general
slice deformation in $y$ satisfying (see the Appendix A)
\begin{equation}
\partial_\mu\partial^\mu\,\hat{\xi}\,^5(x)=
\frac{(1+\gamma)^{-1}}{6\, M_*^3}\,\Big[\frac{S_\lambda^\lambda(x)}{2}
+\ell\,T_{55}(0)-\frac{1+\gamma}{1-\gamma}\,\ell\,\int_0^{y_m}\,
dy\,\partial^\mu T_{\mu 5}\,\Big]\,.\label{mumuxi5}
\end{equation}
Here $y_m$ is the width of matter distribution transverse to the brane, where
$y<y_m$. In terms of the flat-space Green function $\square(x,x')$, this
determines a brane-shift function of the form, with
$T_{55}(0)=\partial^\mu T_{\mu 5}=0$,
\begin{equation}
\hat{\xi}\,^5(x)=\frac{(1+\gamma)^{-1}}{12\, M_*^3}\,
\int d^4 x'\, \square_4 (x,x')\, S_{\lambda}^{\lambda} (x')\,.
\label{braneshift}
\end{equation}
We note that in terms of the $5d$ Neumann Green function $\G_5(X; x',0)$,
the perturbation that follows from~(\ref{TTlinear-in-y'}) is given by
\begin{equation}
h_{\mu\nu}^\p (X)=\bar{h}^\p_{\mu\nu}(X)=-\frac{1}
{2\, M_*^3}\,\int\,d^4x'\,\sqrt{-g}\,\G_5(X; x',0)\,
\Big[ S_{\mu\nu}(x')-\frac{1}{3}\Big(\eta_{\mu\nu}
-\frac{\partial_\mu\,\partial_\nu}{\partial_\lambda^2}\Big) S(x')\Big]\,.
\label{solution-for-h'}
\end{equation}
The first equality follows from $h^\p=0$. Since the brane is located at
$y^\p=0$ ($y=-\hat{\xi}\,^5$), this is the perturbation appropriate to
discuss the observation in the bulk.
Expressing $h_{\mu\nu}$ in terms of Fourier transform w.r.t. the 
brane-world coordinates
\begin{equation}
h_{\mu\nu}(q, y)=\int\,d^4x\,e^{-i\,q_\lambda\,x^\lambda}\,h_{\mu\nu}(x,y)\,,
\end{equation}
we can write $h_{\mu\nu}^\p$ (i.e., using Eq.~\ref{greengen}) in the form,
replacing $M_*^{-3}$ by $16 \pi G_{4+1}$,
\begin{eqnarray}
&&h^\p_{\mu\nu}(q, y)=-8\pi G_{4+1}\,(1-\gamma)^{-1}\,e^{2|y|/\ell}\,
\Big[S_{\mu\nu}(q)-\frac{1}{3}\big(\eta_{\mu\nu}-\frac{q_\mu\,q_\nu}{q^2}
\big)\,S(q)\Big]\times~~~~~~~~~~~~~~\nn\\
&&~~~~~~~~~~~~~~~~~~~~~~~~~~~~~~~~~~~~~~~~~~~~~~~~~
\frac{1}{q}\,\bigg[\frac{H^{(1)}_2(q\ell\,e^{|y|/\ell})}{H^{(1)}_1(q\ell)
+\chi\, q\ell H^{(1)}_2(q\ell)}\bigg]
\end{eqnarray}
In fact, the last term in the first parenthesis can be eliminated using the
gauge transformation induced by $\beta_\mu(x)$ (Eq.~(\ref{hmunutrans})). For
both arguments on the brane, the result is 
\begin{eqnarray}
h_{\mu\nu}(x)&=&-16\,\pi\,G_4\,\int d^4x^\p\,\square_4(x,x')\,
\Big[S_{\mu\nu}(x')-\half\,\eta_{\mu\nu}\,S_\lambda^\lambda (x')\Big]\nn\\
&&~~~~~-8\,\pi\,G_{4+1}\,(1+\gamma)^{-1}\,\int \,d^4x^\p\,\G_{KK}(x,x')\,
\Big[S_{\mu\nu}(x')-\frac{1}{3}\,\eta_{\mu\nu}\,S_\lambda^\lambda (x')\Big]\,,
\label{exact-hmunu}
\end{eqnarray}
where $G_4=G_{4+1}\,\ell^{-1}(1+\gamma)^{-1}$. This can also be decomposed 
into the part corresponding to the matter field and the part corresponding to 
the wall displacement~\cite{GTT}
\begin{equation}
h_{\mu\nu}(x)=h_{\mu\nu}^{(m)}(x)+\frac{2}{\ell}\,\eta_{\mu\nu}\,
\hat{\xi}\,^5(x)\,, 
\end{equation}
where the matter contribution $h_{\mu\nu}^{(m)}$ is given by 
\begin{equation}
h_{\mu\nu}^{(m)}(x)= -8\,\pi\,G_{4+1}\,
\int \,d^4x^\p\,\G_5(x,\ell;x',\ell)\,
\Big[S_{\mu\nu}(x')-\frac{1}{3}\,\eta_{\mu\nu}\,S_\lambda^\lambda (x')\Big]\,,
\end{equation}
and the brane-shift function $\hat{\xi}\,^5(x)$ is given 
by Eq.~(\ref{braneshift}). With the result~(\ref{exact-hmunu}), the 
Eq.~(\ref{tbrane}) would rise to give
\begin{eqnarray}
{\cal A}\equiv\frac{8\pi G_4}{p^2}\bigg[S_{\mu\nu}S^{\mu\nu}
-\half\, S^2\bigg]\,.
\label{newtbrane}
\end{eqnarray}
This is obviously positive definite and ghost (negative norm state) free,
and represents only the
massless spin-$2$ graviton amplitude observed on the brane.

Now, to discuss observation on the brane, we can also define the metric
deformation in trace-reversing form:
$\bar{h}_{\mu\nu}=h_{\mu\nu} -(1/2)\,\eta_{\mu\nu}\, h$, together with the
gauge transformation induced by $\hat{\xi}\,^5$.
In this gauge, the modulo $4$-dimensional gauge transformation, one finds 
\begin{eqnarray}
\bar{h}_{\mu\nu}(x)&=&-8\,\pi\,G_{4+1}\,\int d^4x'\,
\bigg\{\G_5(x,\ell;x',\ell)\,S_{\mu\nu}(x')\nn\\
&&-\eta_{\mu\nu}\,\Big[\G_5(x,\ell;x',\ell)-(1+\gamma)^{-1}\,\frac{2}{\ell}
\,\G_4(x,x')\Big]\,\frac{S_\lambda^\lambda(x)}{6}\bigg\}
\label{trace-reversed1}\,.
\end{eqnarray}
On using the expression~(\ref{fullgreen}), the zero-mode piece
cancels in the term multiplying $S_\lambda^\lambda$. Writing the results in
terms of the $4$-dimensional graviton and KK mode propagators, we find 
\begin{eqnarray}
&&\bar{h}_{\mu\nu}(x)= -16\,\pi\,G_4
\,\int d^4 x'\,\square_4(x, x')\,S_{\mu\nu}(x')\nn\\
&& ~~~~~~~~~~~~~~ -8\,\pi\,G_{4+1}\,(1+\gamma)^{-1}\int d^4x'\, \G_{KK}(x,x')
\Big[S_{\mu\nu}(x')-\frac{1}{6}\,\eta_{\mu\nu}\,S_\lambda^\lambda(x')\Big]
\label{trace-reversed2}\,.
\end{eqnarray}
The first term is the standard result of the four-dimensional gravity, 
with the Planck mass given by
$M_{pl}^2= (1+\gamma)\,M_*^3\,\ell=(16\pi G_4)^{-1}$,
and the second term involves correction from the KK kernel. It appears
that there is no real ghost(negative norm) state even with the GB term. 
Notice that, in the presence of the GB term, the true $4d$ graviton 
propagator on the brane is $\Delta_4(x, x')
=(1+2\chi)^{-1}\,\square_4(x, x')$. However, the 
correction term $(1+2\chi)^{-1}$ to the flat-space $4d$ Green function on
the brane, arised from the gravitational interactions between the matter 
fields living on the brane and the higher curvature terms, when 
multiplied with the term $(1-\gamma)^{-1}$ coming from matching equation 
would rise to give the factor $(1+\gamma)^{-1}$. This is precisely the 
constant factor by which the $4d$ Newton's constant on the brane is 
renormalized in the presence of Gauss-Bonnet term. 
%%%%%%%%%%%%%%%%%%%%%%%%%%%%%%%%%%%%%%%%%%%%%%%%%%%%%%%%%%%%%%%%%
\subsection{Static potential: On and off-brane profile}
%%%%%%%%%%%%%%%%%%%%%%%%%%%%%%%%%%%%%%%%%%%%%%%%%%%%%%%%%%%%%%%%%%
To understand the limiting behavior of the graviton propagator 
at long distances along the brane, one can consider a static point source
at $x'=0$. In the limit $q\ell<<1\,(i.e., |x-x'|=|x|>>\ell)$, the static
potential due to a point source on the brane is given by
\begin{eqnarray}
U(r)&=&\int dt\, \G_5\,(r,\ell;0,\ell)\nn\\
&\simeq& (1-\gamma)^{-1}(1+2\chi)^{-1}\int\frac{d^3p}{(2\pi)^3}\,
e^{ipr}\bigg[-\frac{2}{\ell}\,\frac{1}{p^2}
+\frac{1}{1+2\chi}\,\ell\,\ln(ip\ell/2)\bigg]\nn\\
&\simeq&-(1+\gamma)^{-1}\frac{1}{2\pi r\ell}\bigg[1+
\frac{1-\gamma}{1+\gamma}\,\frac{\ell^2}{2r^2}
+\cdots\bigg]\,.\label{onbranepoten}
\end{eqnarray}
One could obtain the similar 
result from the mode sums, see, for example, Ref.~\cite{IPN1}. 

Now consider the source of gravitational field on the 
brane, $z'=\ell$ ( and $x'=0$). Then for the off-brane graviton
propagator, $z>>\ell$, with the assumption that $r=|x|>>\ell$ and
$z>1/q\gtrsim r$, one can still make the small argument expansion of
$q\ell$ in Eq.~(\ref{greengen}) and obtain
\begin{eqnarray}
&&\G_5(x,z;0,\ell)\simeq(1+\gamma)^{-1}\,\frac{z^2}{\ell^2}
\,\int\frac{d^4p}{(2\pi)^4}\,e^{ipx}\,
\frac{i\pi}{2}\,\ell H_2^{(1)}(qz)\times\label{breakline}\nn\\
&&~~~~~~~~~~~~~~~~~~~~~~~~~~~
\bigg[1+\frac{1-\gamma}{1+\gamma}\,\Big(\ln(q\ell/2+(\Gamma-1/2)
-\frac{\gamma}{1-\gamma}\Big)
(q\ell)^2\bigg]\,.\label{offbrane5}
\end{eqnarray}
In the limit $q\ell<<1$, the static off-brane potential following
from~(\ref{offbrane5}), by integrating over time, is
\begin{equation}
U(r,z)= -(1+\gamma)^{-1}\,\frac{3}{4\pi}\,\frac{1}{\ell z}\,
\bigg(1+\frac{2r^2}{3z^2}\bigg)\bigg(1+\frac{r^2}{z^2}\bigg)^{-3/2}\,
\bigg[1+{\cal O}\big(\ell^2/z^2,\,\ell^2/r^2\big)\bigg]\,.
\label{offbranepro}
\end{equation}
Since $U(r)$ is just proportional to the Green function, this explains the
large $r$ and large $z$ dependence of the propagator off the brane. This
implies that, for a static source and far from the brane, the perturbation
still falls as $h\sim 1/z$. 
%%%%%%%%%%%%%%%%%%%%%%%%%%%%%%%%%%%%%%%%%%%%%%%%%%%%%%%%%%%%
\subsection{The Newtonian limit}
Consider the case of a static and spherically symmetric point source 
of mass $m_*$ localized on the brane, at $\vec{x}=0$. 
The stress tensor can be defined by
\begin{equation}
T_{\mu\nu}=\,m_*\,\delta_{\mu 0}\,
\delta_{\nu 0}\,\delta^{(3)}(x)\,\delta(z)\,.
\end{equation}
With this, Eq.~(\ref{exact-hmunu}) would rise to give
\begin{equation}
h_{00}(x)=\frac{2G_4 m_*}{r}\bigg[1+\frac{2(1-\gamma)}{3(1+\gamma)}\,
\frac{\ell^2}{r^2}\bigg],\, ~~~
h_{ij}(x)=\frac{2G_4 m_*}{r}\bigg[1+\frac{(1-\gamma)}{3(1+\gamma)}\,
\frac{\ell^2}{r^2}\bigg]\,\delta_{ij}\,.
\label{hmunufinal}
\end{equation}
The result is obvious, for the GB term we introduced as
higher-curvature correction still behaves as a ghost-free combination and
hence just renormalizes the Newton constant on the brane, though the
correction terms from the KK mode are different. 
The latter is generic to the RS type higher-dimensional brane 
models~\cite{GTT}, independently whether one has introduced GB 
interaction term or not.

Finally, in the trace-reversed form the metric
deformation~(\ref{trace-reversed2}) gives the expressions
\begin{equation}
\bar{h}_{00}=\frac{4G_4 m_*}{r}\Big[1+\frac{5(1-\gamma)}{12(1+\gamma)}\,
\frac{\ell^2}{r^2}\Big],
\,\,\, \bar{h}_{ij}=\frac{4G_4 m_*}{r}\Big[{\cal O}(0)
+\frac{1-\gamma}{12(1+\gamma)}
\,\frac{\ell^2}{r^2}\Big]\,\delta_{ij}\,.\label{hmunuinTR}
\end{equation}
Thus, only a detailed investigation of the astrophysical implications 
of these results could tell whether or not the RS-type brane model 
is compatible with the cosmological observations.
%%%%%%%%%%%%%%%%%%%%%%%%%%%%%%%%%%%%%%%%%%%%%%%%%%%%%%%%%%%%%%%%
\section{Geodesics in the brane background}
Naively, there could be two possibilities for behavior of a test
particle on the brane, i.e., the test particle is (i) free to move in the
fifth dimension (a geodesic in $AdS_5$ space) or (ii) constrained to move
along the brane by some non-gravitational mechanism
(a geodesic on the brane). Some aspects of geodesics in the RS
(and alternative) brane backgrounds, and in a non-compact $5d$ vacuum
manifold have been studied in~\cite{MUCK} and~\cite{SSW}. Here
we shall discuss more on it in the framework of the RS $3$-brane 
with an extra infinite dimension.

For the $5d$ line element of the form~(\ref{genmetric}), one can write the
Lagrangian, with $\rho(y)\equiv e^{2(c-|y|)/\ell}$, as
\begin{equation}
{\cal L}= \half(ds/d\tau)^2=\half\big[ e^{2(c-|y|)/\ell} g_{\mu\nu}\,
v^\mu\,v^\nu - \dot{y}\,^2\big]\,.\label{lagrangian}
\end{equation}
Here $v^\mu\equiv dx^\mu/d\tau$ is a constant four-vector, $\tau$ is an
affine parameter and the dot represents differentiation w.r. to $\tau$.
For a flat Minkowski $3$-brane, one can set $c=0$.
The solutions to Euler-Lagrange equations for the
Lagrangian~(\ref{lagrangian}) give the following geodesic equations
\begin{equation}
\ddot{x}^\mu={}^{(\tau)}{\sl a}^\mu=-2\,\ell^{-1}\,\partial_y|y|\,
{\dot y}\, v^\mu,~~~ \ddot{y}=- \ell^{-1}\,\partial_y|y|\,e^{2(c-|y|)/\ell}
\,v^2\,,\label{4and5accel}
\end{equation}
where ${}^{(\tau)}{\sl a}^\mu$ is the fourth component of the
$5$-acceleration and $v^2=\eta_{\mu\nu} v^\mu v^\nu$. For an
affine parameter $\tau$ along the path, the first
equation implies a velocity dependent force, while the second
equation shows that a test particle, in general, accelerates in the space
transverse to the brane. These aspects of the fifth force have been
discussed in the literature~\cite{YMC} but in different contexts.

For a non-compact fifth space, it is conceivable that free massive particles
may not move along the brane only, but in general can accelerate in the fifth
dimension, which could be generic in an $AdS$ bulk background, and also 
matter energy may leak 
from the $3$-brane into the bulk. In particular, a highly energetic ordinary
matter can leave the brane and propagate in the $AdS$ bulk. A specific example
of this behavior was noted in~\cite{Ruth} by considering a decay of a
particle of mass $2m$ residing on the brane into two particles of mass $m$.
In the RS scenario, since the massive KK gravitons weakly interact
with the
brane matter, a pair creation of such particles could lead to the transfer of
energy from brane to the bulk. It has been argued in~\cite{Ruth} that for a
meaningful physics these 
particles should behave as dark matter particles of fixed mass and exhibit
the usual behaviors in gravitational interactions without violating locality
and the $4d$ Newton's law on the brane-world volume. The ideas
were further extended in~\cite{SLD} by considering bulk fermions and scalars.

In fact, the RS brane is a
gravitating $3d$ submanifold moving in some higher dimensional
space-time to which ordinary matter is trapped, so that the geodesic for 
any massive matter particle $m>>1/\ell$,
e.g., black hole, on the brane can co-accelerate
with the brane~\cite{ACH}. Further, a $3$-brane metric solution
represents only the core region of a smooth domain wall, so that an observer
sitting on the wall experiences no force. But, moving a certain distance
from the wall, the observer begins to feel acceleration towards the
$AdS$ bulk. How can one reconcile these ideas with the above observation?
It appears that the behavior of extra (fifth) force does not survive in a
different parameterization, but rather brings the correct definition of the
proper time into question in conventional $3+1$ gravity.
In the brane-world scenario, this can be explained. The affine
parameters we define in $4d$ and $5d$ are not the same, but could be related
by $dt\sim e^{-(c-|y|)/\ell}\, d\tau$, where the brane
is located at $y=0$. Since $z=\ell\,e^{|y|/\ell}$, we 
find $t=(z/\ell)\,\tau$, this actually tells that the conformal time 
$t$ defined on the hypersurface is equal to the AdS time $\tau$ times 
the factor $z/\ell$. Furthermore, in the $t$- parameterization, one can 
actually show that ${}^{(t)} {\sl a}^\mu =0$. To explain this 
behavior, we consider below the general solutions to the geodesic 
equations.

Suppose that the $5d$
trajectories are null, $v_a v^a=0$ (more precisely,
$v^a \td_a v^b=0$ and $\lambda$ is an $5d$ affine parameter).
With this hypothesis, the null paths imply that
\begin{equation}
\dot{y}^2=e^{2(c-|y|)/\ell}\,v^2,\,\,\,
\ddot{y}=-\ell^{-1}\,\partial_y|y|\,\dot{y}^2\,,\label{nullpath}
\end{equation}
whose general solution is given by
\begin{equation}
\dot{y}= a_{0}\, e^{(c-|y|)/\ell}\,,\label{ydot}
\end{equation}
where $a_{0}$ is an integration constant and its value can be 
fixed from the initial data on the brane, via
\begin{equation}
(\tau -\tau_{0})= a_{0}^{-1}\,\int_{y_{0}}^{y} e^{|\tilde{y}|/\ell}\,
 d\tilde{y}\label{initial}\,,
\end{equation}
where $y_0=y(\tau_0)$. The time-translation symmetry implies that one can
reparameterize using $\tau\to \tilde{\tau} =\tau_0+\tau\,|a_0|^{-1}$.
Since $d\tilde{\tau}/d\tau>1$, this parameterization preserves the
orientation of $5d$ light cones. Defining
$a_0\,|a_0|^{-1}\, = \,\epsilon$, the above set of equations give
\begin{equation}
{}^{(\tau)}{\sl a}^{\mu}=-2\,\epsilon\,\ell^{-1}\,\partial_y|y|\,
e^{-(c-|y|)/\ell}\, v^{\mu}, ~~~
{} e^{4(c-|y|)/\ell}\, v^2=\epsilon^2\,.\label{epsilon2}
\end{equation}
One can normalize using $\epsilon =0$ for $a_0=0$, and
$\epsilon=\pm 1$ for $a_0\neq 0$. Then from Eq.~(\ref{initial}), 
$\epsilon=0$ implies that $y=y_0$ for all $\tau$, which may 
suggest that for massless particles localized on the brane, e.g., 
photons, there is no motion in the fifth dimension.

Now we perform a parameter transformation $t\to t(\tau)$. From the first
equation of~(\ref{4and5accel}), with $x^{\mu}=x^\mu(t)$, we get
\begin{equation}
{}^{(t)}{\sl a}^{\mu}= -\bigg(\frac{d\tau}{dt}\bigg)^2\,
\bigg(\frac{d^2t}{d\tau^2}+2\ell^{-1}\,\partial_y|y|\,\frac{dy}{d\tau}\,
\frac{dt}{d\tau}\bigg)\,u^\mu\,,\label{newparameter}
\end{equation}
where $u^\mu\equiv dx^\mu/dt$ and $v^\mu=u^\mu\,dt/d\tau$.
Eq.~(\ref{newparameter}) gives ${}^{(t)}{\sl a}^\mu=0$, the standard
geodesic equation for the metric $g_{\mu\nu}(x^\lambda)$, provided that
$dt/{d\tau}=a_1\,e^{-2(c-|y|)/\ell}$. We set $a_1=1$. Then from the second
equation of~(\ref{epsilon2}),
$u_\lambda\,u^\lambda=\epsilon^2$. This implies that the $4d$ geodesic
$u^\mu(t)$ can be time-like ($\epsilon=\pm 1$) or null ($\epsilon=0$), and
the massive matter particles that escape to the bulk could follow the
$5d$ null geodesic. It might be interesting to know whether these particles 
can follow the $5d$ null geodesic and can behave as dark matter particles 
of fixed mass in the bulk. 

In the RS background (a flat Minkowski $3$-brane), the solution to 
the geodesic equation reads 
\begin{equation}
{\bf x}(t)=0\,,~~  y(t)\sim\ell\,\log\big(\epsilon\,t/\ell\big)
\equiv \frac{\ell}{2}\,\log\left(1+t^2/\ell^2\right)\,,
\label{5dlength}
\end{equation}
where $\epsilon =\pm 1$. One then has to consider time $t>0$ once the
matter source is in the bulk. One can
call $t$ the proper time on the brane and $y$ the proper distance 
off the brane. Eq.~(\ref{5dlength}) implies that a large
change in the $4d$ proper time $t$ is accompanied by small changes in the
bulk space ($y$). Working on the $y>0$ side of the brane, we set
$\epsilon =1$. A remarkable result from Eq.~(\ref{5dlength}) is that 
a test particle can reach $y=\infty$ at infinite time $t$, but finite 
proper time $\eta=\pi \ell/2$~\cite{SLD}.

Finally, we note that transverse to the brane the metric 
deformation behaves as $h_{00}\sim m/M_*^2\times 1/z$ and the 
horizon size for a static black-hole grows like $r_h\sim z_h$. 
Therefore, for a black hole of mass $m$ on the brane, if the 
horizon size along the brane grows like $\sim m$, the thickness 
transverse to the brane grows only like $\sim \log m$, i.e.,
\begin{equation}
y_h\sim \ell\, \log\bigg(\frac{m}{M_*^{2}}\,\frac{1}{\ell}\bigg)\,.
\end{equation}
This entails a panecake shape of the black hole and has been adequately 
discussed in the literature~\cite{CHR}.
%%%%%%%%%%%%%%%%%%%%%%%%%%%%%%%%%%%%%%%%%%%%%%%%%%%%%%%%%%%%%%%%%%%%%%%%%%
\section{Conclusion}
%%%%%%%%%%%%%%%%%%%%%%%%%%%%%%%%%%%%%%%%%%%%%%%%%%%%%%%%%%%%%%%%%%%%%%%%%
A linearized treatment of Einstein-Gauss-Bonnet (EGB) gravity in the
presence of a singular positive tension $3$-brane and localized matter
distribution is presented. The full graviton propagator is shown to be
well behaved in all distance scales and hence on the brane one can still
reproduce the zero-mode solution as a localized gravity with correct 
momentum and tensor structures even in the presence of GB 
interaction term. In a linearized analysis, this paper has 
outlined the ghost problem (negative norm state) of the GB term 
in the brane background and ways to resolve it and many other 
interesting features of EGB gravity and gravitational potential 
corrections with GB interaction. It is shown that for the matter 
localized $3$-brane in an $AdS_5$, a test particle on the brane 
can still follow the time like geodesic if it is null in five space.
\section*{Acknowledgements}
We would like to thank J. Erlich, E. Katz, A. Karch, Z. Kakushadze,
H. M. Lee, S.-H. Moon, Y. S. Myung and S. S. Seahra for useful 
discussions and correspondences. This work was supported in part 
by BK-21 Initiative in Physics, SNU Project 2 and the Natural Sciences 
and Engineering Research Council of Canada.
%%%%%%%%%%%%%%%%%%%%%%%%%%%%%%%%%%%%%%%%%%%%%%%%%%%%%%%%%%%%%%%%%%%%%%%
%\section*{Appendix}
\renewcommand{\thesection}{\mbox{\Alph{section}}}
\renewcommand{\theequation}{\mbox{\Alph{section}.\arabic{equation}}}
\setcounter{section}{0}
\setcounter{equation}{0}
\section{Appendix A. Linearized equations and boundary solutions}
Consider the metric fluctuations in the form
$g_{ab}= e^{-2A(z)}\big(\eta_{ab}+h_{ab}\big)$. Then the linearized equations
for the action~(\ref{action}) with $d+1=5$, in
axial gauge $h_{a 5}=0$, take the form
%%%%%%%%%%%%%%%%%%%%%%%%%%%%%%%%%%%%%%%%%%%%%%%%%%%%%%%%%%%%%%
\begin{eqnarray}
&&\left[-\big(1+4\alpha\kappa e^{2A} A^{\p\p}\big)\partial_\lambda^2
-\big(1-4\alpha\kappa e^{2A} {A^\p}\,^2\big)\partial_z^2
+ 3\,A^\p  \partial_z\right]\big(h_{\mu\nu}-\eta_{\mu\nu} h\big)\nn\\
&&+\left(1-4\alpha\kappa e^{2A}
\big(2{A^\p}\,^2-A^{\p\p}\big)\right) 
\left(2\partial_{(\mu}\partial^\lambda h_{\nu)\lambda}
-\partial_\mu \partial_\nu h \right) 
- \big(1+4\alpha\kappa e^{2A} A^{\p\p}\big)\times\nn\\
&&\eta_{\mu\nu}\partial_\lambda\partial_\rho
h^{\lambda\rho} +4\alpha\kappa e^{2A} A^\p
\left[\big(2A^{\p\p}-{A^\p}\,^2\big) \partial_zh_{\mu\nu}
+\big(A^{\p\p}-2{A^\p}\,^2\big)\eta_{\mu\nu} \partial_zh\right]
=\kappa\, T^{(m)}_{\mu\nu}\,,\\
&& ~~~~~~~~~~~~~~~~~~~~~~~~~~~~~~~~~~~~~~~~~
\big(1-4\alpha\kappa e^{2A}{A^\p}\,^2\big)\,
\partial_z \big(\partial^\lambda h_{\mu\lambda}-\partial_\mu h\big)
=\kappa\, T^{(m)}_{5\mu},\label{aymucomp}\\
&& ~~~~~~~~~~~~~~~~~~~~~-\big(1-4\alpha\kappa e^{2A}{A^\p}\,^2\big)\,
\left(\partial^\mu\partial^\nu (h_{\mu\nu}
-\eta_{\mu\nu}h)+3 A^\p\partial_z h\right)=\kappa\, T^{(m)}_{55}.
\label{ayycomp}
\end{eqnarray}
With a coordinate transformation $|y|=\ell\,\log(|z|/\ell+1)$, or
defining the background metric in $y$-coordinate~(\ref{RSbgmetric}), the
above set of equations take the form
\begin{eqnarray}
&&\big(1-4\alpha\kappa{A^\p}\,^2+4\alpha\kappa A^{\p\p}\big)\,
e^{2|y|/\ell}\,\left(\partial_\lambda^2(h_{\mu\nu}-\eta_{\mu\nu}h)+
\partial_\mu\partial_\nu h-
2\partial_{(\mu}\partial^\lambda h_{\nu)\lambda}+
\eta_{\mu\nu}\partial_\lambda\partial_\rho h^{\lambda\rho}\right)\nn\\
&&+\left(\partial_y^2-4\alpha\kappa\,\big({A^\p}\,^2\,
\partial_{y}^2+2A^\p A^{\p\p}\,\partial_y\big)
-4{A^\p}\,^2\big(1-4\alpha\kappa {A^\p}\,^2\big)
+2 A^{\p\p}\big(1-12\alpha\kappa {A^\p}\,^2\big)\right)\times\nn\\
&& ~~~~~~~~~~~~~~~~~~~~~~~~~~~~~~~~~~~~~~~~~~~~~~~~~~~~~~~~~~~~~~~~~~~~~~~
(h_{\mu\nu}-\eta_{\mu\nu} h)=-\kappa\,T_{\mu\nu}^{(m)}\,
\label{munucompo-in-y}\\
&&~~~~~~~~~~~~~~~~~~~~~~~~~~~~~~~~~~~~~
\big(1-4\alpha\kappa {A^\p}\,^2\big)\,
\partial_y \left(e^{2|y|/\ell}(\partial_\mu h-\partial^\lambda
h_{\mu\lambda}\big)\right)=-\kappa\, T^{(m)}_{5\mu},\label{munucomp-in-y}\\
&&~\big(1-4\alpha\kappa {A^\p}\,^2\big)\,
e^{2|y|/\ell}\,\left(e^{2|y|/\ell}\,\partial^\mu\partial^\nu (h_{\mu\nu}
-\eta_{\mu\nu}h)+3 A^\p(\partial_y h+2\,\ell^{-1} h)\right)
=-\kappa\, T^{(m)}_{55}\,,\label{55compo-in-y}
\end{eqnarray}
where $A^\p=\ell^{-1}\partial_y|y|$, $A^{\p\p}=2\,\ell^{-1}\delta(y)$ 
and ${A^\p}^2 A^{\p\p}=(2/3\ell^2)\,\delta(y)$. 
In the presence of matter source on the brane, one has to put no
restriction to the $4d$ non-TT components of metric fluctuations.
Working on the $y>0$ ($z>\ell$) side of the
brane (i.e. $A^{\p\p}=0$), taking the trace of~(\ref{munucompo-in-y})
and subtracting the Eq.~(\ref{55compo-in-y}) from the resulting expression,
we get
\begin{equation}
\partial_y(2\ell^{-1}+\partial_y )\,h=\frac{(1-\gamma)^{-1}}{3\,M_*^3}
\Big[\,T_\mu^\mu-2e^{-2y/\ell}\,T_{55}\,\Big]\,.
\label{bulk-equation}
\end{equation}
Conservation of $T$ in the bulk brings this in the form
\begin{equation}
\partial_y\Big[\big(\partial_y + 2\,\ell^{-1}\big)\,h + \frac{(1-\gamma)^{-1}\,
\ell}{3\,M_*^3}\,e^{-2y/\ell}\,T_{55}(y)\Big]
=\frac{(1-\gamma)^{-1}\,\ell}{3\,M_*^3}\partial_\mu T^{\mu 5}\,.
\label{Tconservation}
\end{equation}
This can be solved (integrated) with initial boundary condition of the first
bracket term supplied by the trace of~(\ref{munucompo-in-y}). In particular, 
with GB term, one first needs to address a subtlety in the
boundary condition, because of the terms 
like $sgn(y)\delta(y)\,\partial_y h$ and $sgn(y)^2\,\delta(y)$. However, one
can unambiguously fix them by properly regularizing the $\delta$-function as 
we did in the main text. 

Given these considerations, taking the trace of~(\ref{munucompo-in-y}) and
integrating from $\epsilon_-$ to $\epsilon_+$, in the limit $\epsilon\to 0$,
we arrive at
\begin{equation}
(1-\gamma) (\partial_y + 2\ell^{-1})\,h|_{0_+}-\frac{2}{3}\,\gamma\,\ell\,
\big(\partial_\lambda\partial_\rho h^{\lambda\rho}-\partial_\lambda^2 h\big)
|_{0_+}=\frac{1}{6M_*^3}\,S_\mu^\mu\,.\label{boundary-condition}
\end{equation}
But from the $(55)$-component of the fluctuations, Eq.~(\ref{55compo-in-y}), 
we have
\begin{equation}
\big(\partial_\lambda\partial_\rho h^{\lambda\rho}-\partial_\lambda^2h\big)|
_{0_+}=-3\ell^{-1}(\partial_y+2\,\ell^{-1})\,h|_{0_+}-M_*^{-3}\,
(1-\gamma)^{-1} T_{55}(0)\,.
\end{equation}
Substituting this back into~(\ref{boundary-condition}), we get
\begin{equation}
(\partial_y+2\ell^{-1})h|_{0_+}=\frac{(1+\gamma)^{-1}}{3M_*^3}\,
\left[\frac{S_\mu^\mu}{2}-\frac{2\gamma\,\ell}{1-\gamma}\, T_{55}(0)\,\right]
\,.\label{initial-condi}
\end{equation}
The trace $h$ involves a growing component transverse
to the brane, and this may lead to failure of the linear approximation when
$S_\mu^\mu\neq 0$. However,
one can eliminate such growth in $h$ from the initial
condition~(\ref{initial-condi}) by a coordinate transformation of the
form~(\ref{hmunutrans}), and then we may integrate Eq.~(\ref{Tconservation})
to eliminate resultant growth in $h$ with a gauge choice
\begin{equation}
\partial_\mu\partial^\mu\,\hat{\xi}\,^5(x)=
\frac{(1+\gamma)^{-1}}{6\, M_*^3}\,\left[\frac{S_\mu^\mu(x)}{2}
+\ell\,
T_{55}(0)-\frac{1+\gamma}{1-\gamma}\,\ell\,\int_0^{y_m}\,
dy\,\partial^\mu T_{\mu 5}\,\right]\,.\label{gauge-shift}
\end{equation}
%%%%%%%%%%%%%%%%%%%%%%%%%%%%%%%%%%%%%%%%%%%%%%%%%%%%%%%%%%%%%%%%%%%%%%
The gauge shift induced by~(\ref{gauge-shift}) can be used to determine the
boundary conditions for $h^\p_{\mu\nu}$ and hence to find the form of
the source term $\Sigma_{\mu\nu}(x')$. The boundary condition on
$h_{\mu\nu}$ at the brane is readily determined by integrating
Eq.~(\ref{munucompo-in-y}) from just below to just above the brane (i.e.,
Israel junction condition) and enforcing symmetry under $y\to -y$, which is
read as
\begin{eqnarray}
&&\gamma\,\ell\, e^{2|y|/\ell}\,\big(\partial_\lambda^2 h_{\mu\nu}
-\eta_{\mu\nu}\,\partial_\lambda^2 h+\partial_\mu\partial_\nu h-
2\partial_{(\mu}\partial^\lambda h_{\nu)\lambda}+
\eta_{\mu\nu}\partial^\lambda\partial^\rho h_{\lambda\rho}\big)_{|{y=0_+}}~~~~~~~~~~~\nn\\
&&~~~~~~~~~~~~~~~~~~~~~~
+(1-\gamma)\, (\partial_y+2\,\ell^{-1})\,
(h_{\mu\nu}-\eta_{\mu\nu} h)_{|{y=0_+}}
=-\frac{\kappa}{2}\,S_{\mu\nu}\,.
\end{eqnarray}
In terms of the $h^\p_{\mu\nu}$, i.e., under the gauge
transformation in the fluctuation $h_{\mu\nu}$, Eq.~(\ref{hmunutrans}),
this becomes
\begin{eqnarray}
&&\gamma\,\ell\, e^{2|y|/\ell}\,\Big(\partial_\lambda^2 h^\p_{\mu\nu}
-\eta_{\mu\nu}\partial_\lambda^2 h^\p
+\partial_\mu\partial_\nu h^\p- 2\partial_{(\mu}\partial^\lambda h^\p_{\nu)
\lambda}+\eta_{\mu\nu}\partial^\lambda\partial^\rho 
h^\p_{\lambda\rho}\Big)_{|y=0_+}\nn\\
&&~~~~~~~~~~~~~~~~~~~~~~+(1-\gamma)\,(\partial_y + 2\,\ell^{-1})\,
(h^\p_{\mu\nu}-\eta_{\mu\nu} h^\p)_{|y=0_+}\nn \\
&&~~~~~~~~~~=-\frac{\kappa}{2}\,S_{\mu\nu}\,-4\gamma (\partial_\mu\partial_\nu
-\eta_{\mu\nu}\partial_\lambda^2)\,\hat{\xi}\,^5
-2(1-\gamma)\,(\partial_\mu\partial_\nu
-\eta_{\mu\nu}\partial_\lambda^2)\,\hat{\xi}\,^5\nn \\
&&~~~~~~~~~= - \frac{\kappa}{2}\,S_{\mu\nu}
-2(1+\gamma)\,(\partial_\mu\partial_\nu
-\eta_{\mu\nu}\partial_\lambda^2)\hat{\xi}\,^5\,.
\label{finalb.c}
\end{eqnarray}
One obtains exactly the same boundary condition for $\bar{h}^\p_{\mu\nu}$,
justifying the gauge choice $\bar{h}^\p=0$. The tracefree gauge $h^\p = 0 $
generates an extra
unphysical scalar degree of freedom which may correspond to the graviscalar
mode. This extra polarization state, however, can be
compensated by shifting the brane to $y^\p=0\,(y=-\xi^5(x))$.
The cancellation of the unphysical scalar mode is, indeed, needed for the
validity of linearized approximation.
Further a choice $\partial^{\mu} h^\p_{\mu\nu} = 0$ ensures the conservation
of the energy momentum tensor on the brane and hence the $4d$ general
covariance of the $3$-brane world volume. With the gauge choice
$h^\p=\partial^\mu h^\p_{\mu\nu}=0$, (\ref{finalb.c}) one ends up with the
correct boundary condition for $h^\p_{\mu\nu}$.

Now, using~(\ref{gauge-shift}) and considering the case
$T_{55}(0)=\partial^\mu T_{\mu 5}=0$, the
solution for $h^\p_{\mu\nu}$, in terms of the $5d$ Neumann Green function,
is given by
\begin{equation}
h_{\mu\nu}^\p (X)=\bar{h}^\p_{\mu\nu}(X)=-\frac{1}
{2\, M_*^3}\,\int\,d^4x'\,\sqrt{-g}\,G_5(X; x',0)\,\Sigma_{\mu\nu}(x')\,,
\label{sol-hprime}
\end{equation}
where the source term is given by
\begin{equation}
\Sigma_{\mu\nu}(x')= S_{\mu\nu}(x')-\frac{1}{3}\Big(\eta_{\mu\nu}
-\frac{\partial_\mu\,\partial_\nu}{\partial_\lambda^2}\Big) S(x')\,.
\end{equation}
Obviously, $\Sigma_{\mu\nu}(x')$ is transverse and trace-free, and
includes the contribution of the brane-shift function, which does play a
role of the source in the RS gauge.
%%%%%%%%%%%%%%%%%%%%%%%% reference %%%%%%%%%%%%%%%%%%%%%%%%%%%%%%%


\begin{thebibliography}{100}

\bibitem{ADD} N. Arkani-Hamed, S. Dimopoulos and G. Dvali, Phys. Lett.
{\bf B429}, 263 (1998); I. Antoniadis, N. Arkani-Hamed,
S. Dimopoulos and G. Dvali, Phys. Lett. {\bf B436} (1998) 257.

\bibitem{RS} L. Randall and R. Sundrum, Phys. Rev. Lett. {\bf 83} (1999)
3370, [hep-ph/9905221].

\bibitem{CBK} A. G. Cohen and D. B. Kaplan, Phys. Lett. {\bf B470} (1999)
52; M. Goberashvili, hep-ph/9812296.

\bibitem{MRK} Y. Kim, S.-H. Moon and S.-J. Rey, Nucl. Phys. {\bf B602} 
(2001) 467, hep-th/0012165.

\bibitem{RS1} L. Randall and R. Sundrum, Phys. Rev. Lett. {\bf 83} (1999)
4690, [hep-th/9906064].
\bibitem{VAR} V. A. Rubakov and M. E. Shaposhnikov. Phys. Lett. {\bf B125}
(1983) 136; K. Akama, Lect. Notes Phys. {\bf 176} (1982) 267,
[hep-th/0001113]; M. Visser, hep-th/9910093.
%%%%%%%%% Cosmology %%%%%%%%%%%%%%%%%%%%%%%%%%%%%%%%

%\bibitem{CGKT} C. Csaki, M. Graesser, C. Kolda and J. Terning,
%Phys. Lett. {\bf B462} (1999) 34; J. M. Cline, C. Grojean and G. Servant,
%Phys. Rev. Lett. {\bf 83} (1999) 4245;
%H. B. Kim and H. D. Kim, Phys. Rev. {\bf D61} (2000) 064003; P. Kanti,
%I. I. Kogan, K. A. Olive and M. Pospelov, Phys. Lett. {\bf B468} (1999) 31;
%P. Binetruy, C. Deffayet and U. Ellwanger, Phys. Lett. {\bf B477} (2000)
%285.
%%%%%%%%%%%%%%% Black Hole %%%%%%%%%%%%%%%%%%%%%%%%%%%

\bibitem{CHR} A. Chamblin, S. W. Hawking and H. S. Reall, Phys. Rev.
{\bf D61} (2000) 065007, [hep-th/9909205]; R. Emparan, G. T. Horowitz
and R. C. Myers, JHEP {\bf 0001} (2000) 07, [hep-th/9911043];
S. B. Giddings and E. Katz, hep-th/0009176.
%%%%%%%%%%%%%% SUSY and SUGRA %%%%%%%%%%%%%%%%%%%%

\bibitem{BMC} K. Behrndt and M. Cvetic, Phys. Lett. {\bf B475} (2000) 253;
A. Chamblin and G. W. Gibbon, Phys. Rev. Lett. {\bf 84} (2000) 1090,
[hep-th/9909130]; R. Kallosh and A. Linde, JHEP {\bf 0002} (2000) 005;
D. Youm, Nucl. Phys. {\bf B576} (2000) 106; M. Cvetic, H. Lu and C. N.
Pope, Class. Quant. Grav. {\bf 17} (2000) 4867; {\it ibid},
hep-th/0007209; M. J. Duff, J. T. Liu and K. S. Stelle, hep-th/0007120.
%%%%%%%%%%%%%%%%%%%%%%%%% 5d gravity and string %%%%%%%%%%%%%%%%%%%%%%%

\bibitem{KSS} S. Kachru, M. Schulz and E. Silverstein, Phys. Rev. {\bf D62}
(2000) 045021, [hep-th/0001206]; C. S. Chan, P. L. Paul and H.
Verlinde, Nucl. Phys. {\bf B581} (2000) 156; B. R. Greene, K. Schaim and
G. Shiu, Nucl. Phys. {\bf B584} (2000) 480; S. Nojiri and S. D. Odintsov,
Phys. Rev. {\bf D62} (2000) 064018, [hep-th/9911152].
%%%%%%%%%%%%%%5 AdS Walls, AdS-CFT and Gravity %%%%%%%%%%%%%%%%%

\bibitem{SJR} M. Cvetic, S. Griffies and S.-J. Rey, Nucl. Phys. {\bf B381}
(1992) 301.

\bibitem{PKR} O. Dewolfe, D. Z. Freedman, S. S. Gubser and A. Karch, Phys.
Rev. {\bf D62} (2000) 046008, [hep-th/9909134];
P. Kraus, JHEP {\bf 9912} (1999) 011; S. S. Gubser, hep-th/9912001.
%%%%%%%%%%%%%%%%%%% RG flow  %%%%%%%%%%%%%%%%%%%%%%%%%

%\bibitem{SKT} K. Skenderis and P. K. Townsend, Phys. Lett. {\bf B468}
(1999) 46.
%%%%%%%%%%%%%%%%% Holography %%%%%%%%%%%%%%%%%%%%%%%%%%%%%%

%\bibitem{HVR} H. Verlinde, Nucl. Phys. {\bf B580} (2000) 264,
%[hep-th/9906182], J. de Boer, E. Verlinde and H. Verlinde, JHEP {\bf 0008}
%(2000) 003.

\bibitem{IMV} V. D. Ivashchuk, M. Kenmoku and V. N. Melnikov, Grav. Cosmol. 
{\bf 6} (2000) 225, gr-qc/0101043.

\bibitem{BCM} W. S. Bae, Y. M. Cho and S.-H. Moon, JHEP {\bf 0103} (2001)
039, [hep-th/0012221].

%%\bibitem{HEW} P. Horava, E. Witten, Nucl. Phys. {\bf B460} (1996) 506.
\bibitem{GTT} J. Garriga and T. Tanaka, Phys. Rev. Lett. {\bf 84}, (2000)
2778, [hep-th/9911055].

\bibitem{CEHS}  C. Cs$\acute{a}$ki, J. Erlich, T.J. Hollowood and
Y. Shirman, Nucl. Phys. {\bf B581} (2000) 309, [hep-th/0001033].
\bibitem{GKR} S. B. Giddings, E. Katz and L. Randall, JHEP {\bf 0003} (2000)
23, [hep-th/0002091].

%\bibitem{KAN} G. Kang and H.W. Lee and Y. S. Myung, Phys. Lett.
%{\bf B478} (2000) 294, [hep-th/0001107], M. G. Ivanov and I. V. Volovich,
%hep-th/9912242, $v_4$.

\bibitem{IYA} I. Ya. Aref'eva, M. G. Ivanov, W. Mueck, K. S.
Viswanathan and I. V. Volovich, Nucl. Phys. {\bf B590} (2000) 273,
[hep-th/0004114].

\bibitem{YSM} G. Kang and Y. S. Myung, Phys. Lett. {\bf B483} (2000) 235,
[hep-th/0003162]; Y. S. Myung, hep-th/0009117.

\bibitem{KAKU} Z. Kakushadze, Phys. Lett. {\bf B497} (2001) 125,
[hep-th/0008128].

\bibitem{KL} J. E. Kim and H. M. Lee, Nucl. Phys. {\bf B602} (2001) 346, 
hep-th/0010093.

\bibitem{KLR} A. Karch and L. Randall, JHEP {\bf 0105} (2001) 008, 
hep-th/0011156.

\bibitem{Kogan2} I.~I.~Kogan, S.~Mouslopoulos and A.~Papazoglou, Phys. Lett. 
{\bf B501} (2001) 140, hep-th/0011141.

\bibitem{IPN1} I. P. Neupane, Phy. Lett. {\bf B512} (2001) 137, hep-th/0104226.

\bibitem{SMS} T. Shiromizu, K.-I. Maeda and M. Sasaki, Phys. Rev. {\bf D62}
(2000) 024012, [gr-qc/9910076].

\bibitem{ACF} Modern Kaluza-Klein Theories, eds. T. Appelquist,
A. Chodes and P. G. Freud, (Addison and Wesley, 1987).

\bibitem{ZWI} B. Zweiebach, Phys. Lett. {\bf B156} (1985) 315; B. Zumino,
Phys. Rept. {\bf 137} (1986) 109.

\bibitem{KIM} J. E. Kim, B. Kyae and H. M. Lee, Phys. Rev. {\bf D62}
(2000) 045013, hep-ph/9912344.

\bibitem{DTD} N. Deruelle and T. Dolezel, Phys. Rev. {\bf D62} (2000)
103502, [gr-qc/0004021]; T. Dolezel, hep-th/0012160.

\bibitem{NEM} N. E. Mavromatos and J. Rizos, Phys. Rev. {\bf D62} (2000)
124004, hep-th/0008074.

\bibitem{IPN} I. P. Neupane, JHEP {\bf 0009} (2000) 040, [hep-th/0008190].

\bibitem{CZK} O. Corradini and Z. Kakushadze, Phys. Lett. {\bf B494}
(2000) 302, [hep-th/0009022]; Z. Kakushadze, Nucl. Phys. {\bf B589} (2000)
75.

\bibitem{NOO} S. Nojiri and S. D. Odintsov, JHEP {\bf 0007} (2000) 049;
S. Nojiri, S. D. Odintsov and S. Ogushi, hep-th/0010004.

\bibitem{Marek} K.~A.~Meissner and M.~Olechowski, Phys. Rev. Lett. {\bf 86} 
(2001) 3708, hep-th/0009122. 

\bibitem{IPN4} I.~P.~Neupane, hep-th/0106100; K.~A.~Meissner and 
M.~Olechowski, hep-th/0106203.

%\bibitem{BKS} A. Brandhuber and K. Sfetsos, JHEP {\bf 9910} (1999) 013;
%R. Dick and D. Mikulovicz, Phys. Lett. {\bf B476} (2000)
%363, [hep-th/0001013].

%\bibitem{DMV} H. van Dam and M. Veltman, Nucl. Phys. {\bf B22} (1970) 397;
%V. I. Zakharov, JTEP Lett. {\bf 12} (1970) 312: Y. M. Cho and S. W. Zoh,
%Phys. Rev. {\bf D46} (1992) 2290.

\bibitem{CEH} C. Cs$\acute{a}$ki, J. Erlich and T.J. Hollowood, Phys. Rev.
Lett. {\bf 84} (2000) 5932, [hep-th/0002161]; {\it ibid } Phys. Lett.
{\bf B481} (2000) 107, [hep-th/0003020].

\bibitem{DGP} G. Dvali, G. Gabadadze and M. Porrati, Phys. Lett. {\bf B484}
(2000) 112, [hep-th/0002190].

\bibitem{RGV} R. Gregory, V.A. Rubakov and S. M. Sibiryakov, Phys. Rev.
Lett. {\bf 84} (2000) 5932, [hep-th/0002072]; {\it ibid} Phys.
Lett. {\bf B489} (2000) 203, [hep-th/0003045], G. Dvali, G. Gabadadze and
M. Porrati, Phys. Lett. {\bf B484} (2000) 129, [hep-th/0003054];
C. Csaki, J. Erlich, T. J. Hollowood and J. Terning, hep-th/0003076.

\bibitem{MUCK} W. M\"uck, K. S. Viswanathan and I.V. Volovich,
Phys. Rev. {\bf D62} (2000) 105019, [hep-th/0002132].
\bibitem{SSW} S. S. Seahra and P. S. Wesson, to appear in Gen. Rel. Grav.

\bibitem{YMC} Y. M. Cho and D. H. Park, Gen. Rel. Grav. {\bf 23} (1991)
741; P. S. Wesson, B. Mashhoon, H. Liu and W. N. Sajko, Phys. Lett.
{\bf B456} (1999) 34.

\bibitem{Ruth} R. Gregory, V. A. Rubakov and S. M. Sibiryakov, Class. Quant.
Grav. {\bf 17} (2000) 4437, [hep-th/0003109].

\bibitem{SLD} S. L. Dubovsky, V. A. Rubakov and P. G. Tinyakov, Phys. Rev.
{\bf D62} (2000) 105011, [hep-th/0006046].

\bibitem{ACH} A. Chamblin, Clas. Quan. Grav. {\bf 18} (2001) L17,
[hep-th/0011128].
%\end{references}
\end{thebibliography}
\end{document}